# ORACLE: A Real-time, Hierarchical, Deep Learning Photometric Classifier for the LSST

Ved G. Shah[1,2,3,4] , Alex Gagliano[5,6,7] , Konstantin Malanchev[8] , Gautham Narayan[2,3,9,10] ,
Alex I. Malz[8] , and the LSST Dark Energy Science Collaboration

[1] Center for Interdisciplinary Exploration and Research in Astrophysics, Northwestern University, Evanston, IL, USA; vedgs2@illinois.edu
[2] NSF-Simons AI Institute for the Sky (SkAI), 172 E. Chestnut St., Chicago, IL 60611, USA
[3] Department of Astronomy, University of Illinois Urbana-Champaign, Urbana, IL, USA
[4] Siebel School of Computing and Data Science, University of Illinois Urbana-Champaign, Urbana, IL, USA
[5] The NSF AI Institute for Artificial Intelligence and Fundamental Interactions, USA
[6] Center for Astrophysics |Harvard & Smithsonian, Cambridge, MA 02138, USA
[7] Department of Physics, Massachusetts Institute of Technology, Cambridge, MA 02139, USA
[8] McWilliams Center for Cosmology and Astrophysics, Department of Physics, Carnegie Mellon University, Pittsburgh, PA 15213, USA
[9] Center for AstroPhysical Surveys, National Center for Supercomputing Applications, Urbana, IL 61801, USA
[10] Illinois Center for Advanced Studies of the Universe, University of Illinois Urbana-Champaign, Urbana, IL 61801, USA
Received 2024 December 20; revised 2025 August 15; accepted 2025 October 2; published 2025 December 2

## Abstract

We present the Online Ranked Astrophysical CLass Estimator (ORACLE), the first hierarchical deep-learning model for real-time, context-aware classification of transient and variable astrophysical phenomena. ORACLE is a recurrent neural network with gated recurrent units, and has been trained using a custom hierarchical cross-entropy loss function to provide high-confidence classifications along an observationally driven taxonomy with as little as a single photometric observation. Contextual information for each object, including host galaxy photometric redshift, offset, ellipticity, and brightness, is concatenated to the light-curve embedding and used to make a final prediction. Training on ∼0.5M events from the Extended LSST Astronomical Time-series Classification Challenge, we achieve a top-level (transient versus variable) macroaveraged precision of 0.96 using only 1 day of photometric observations after the first detection in addition to contextual information, for each event; this increases to >0.99 once 64 days of the light curve has been obtained, and 0.83 at 1024 days after first detection for 19-way classification (including supernova subtypes, active galactic nuclei, variable stars, microlensing events, and kilonovae). We also compare ORACLE with other state-of-the-art classifiers and report comparable performance for the 19-way classification task, in addition to delivering accurate top-level classifications much earlier. The code and model weights used in this work are publicly available at our associated GitHub repository (https://github.com/uiucsn/Astro-ORACLE).

## 1. Introduction

The chemical composition, density structure, and kinematics of a transient astrophysical phenomenon are all resolved through spectroscopy. For example, spectroscopy reveals the presence of elements in supernova ejecta, changes in density in collapsing stars through the broadening of emission lines, and the motion of transients by measuring redshifts.

The fraction of discovered transient events that can be characterized spectroscopically, given current spectroscopic resources, will drop to <1% with the advent of the Vera C. Rubin Observatory's Legacy Survey of Space and Time (LSST; Ž. Ivezić et al. 2019). LSST's high étendue, the product of the collecting area and solid angle seen by the detector, results in a survey speed an order of magnitude higher than any existing telescope, and it will discover ∼100× more transients, particularly at high ($z \lesssim 1$) redshift.

Our paucity of spectroscopic follow-up resources, balanced against the rapid evolution of some of the rarest phenomena, has driven a need for near-instantaneous triaging to select the objects that will maximize an instrument's scientific return. While spectroscopic analysis reveals the underlying physics of a transient, the most immediate use of spectroscopy is to place an object within a taxonomic context—the realm of classification.

Lacking spectra, many groups have focused on developing photometric classifiers for diverse transient and variable phenomena (e.g., M. Lochner et al. 2016; D. Muthukrishna et al. 2019; A. Möller & T. de Boissière 2020; R. Carrasco-Davis et al. 2021; A. Gagliano et al. 2023; G. Cabrera-Vives et al. 2024; B. M. O. Fraga et al. 2024). The challenges of this approach are manifold: photometry, while rapidly obtained, preserves minimal spectral information. Furthermore, optical transients span a dynamic range of timescales from hours to years, and increasing discovery rates have translated to lower temporal coverage for a single event (300 visits per year in three filters for the Zwicky Transient Facility (ZTF) compared to 100 visits per year in six filters for LSST; M. J. Graham et al. 2019). Consequently, photometric classifiers typically only achieve sufficient performance when considering a small subset of the full transient zoo, and/or only when most of the light curve has been observed.

In the early phases of light curve evolution, significant observational degeneracies exist between classes. As a result, several models in the literature struggle to achieve performance competitive with full-phase classification (D. Muthukrishna et al. 2019; A. Gagliano et al. 2023; G. Cabrera-Vives et al. 2024). However, it is typically more beneficial to determine the object's class early to decide if it merits spectroscopic follow-up. While this problem is difficult to address directly, it may be beneficial to produce more confident, albeit less granular, classifications at early times.

There is a fundamental risk-reward trade-off between obtaining information early with the use of spectroscopic resources, but triggering on commonplace sources; or waiting much longer for a confident photometric classification, but then potentially missing valuable temporally evolving emission from the source (in addition to having to expend more time to obtain spectroscopic follow-up of the fainter transient at a later phase, or missing it entirely).

We argue that accurate photometric triaging requires classifying transients with increasing granularity as more data becomes available. To address this accuracy-latency trade-off, we introduce the Online Ranked Astrophysical CLass Estimator (ORACLE): a real-time, hierarchical photometric classification model for LSST that provides actionable class information at every phase in a transient's evolution.

We structure this paper as follows. In Section 2, we discuss the motivation behind the development of ORACLE. In Sections 3 and 4, we document the data set, taxonomy, and features used in this work. In Section 5, we highlight the recurrent neural network (RNN) architecture that is used for ORACLE. In Section 6, we discuss the mathematical formulation for the custom hierarchical loss functions used to train ORACLE. In Section 7, we report the model hyperparameters and describe our training workflow. In Section 8, we report the performance of ORACLE and compare it with other photometric classifiers from the literature. In Section 10, we provide concluding thoughts and discuss ORACLE's role in the broader context of astrophysical classification. In Section 11, we discuss current and future projects that will make use of the groundwork we have laid here.

## 2. Motivation

Taxonomies cluster diverse astrophysical phenomena into a comprehensible framework as a first attempt at understanding their fundamental nature. Transient and Variable phenomena can be classified in different ways: by the type of temporal variability they exhibit, such as periodic or stochastic, or by the physical processes driving them. For example, supernovae (SNe) are categorized based on their presumed explosion mechanism, such as thermonuclear or core collapse. Because each event is unique, transients can be further subdivided into smaller and smaller bins based on a panoply of observed and intrinsic properties, including their progenitor systems, formation channels, and photometric or spectroscopic characteristics (for example, calcium-rich transients can be separated from SNe Ib/c by the presence of strong Ca emission lines in their spectra). We present a time-domain taxonomy in Figure 1, which we adopt for this work. This taxonomy is loosely inspired by the structure of our training data set (see Section 3), in order to maintain consistency with the training labels. Taxonomies are rarely complete, and typically do not consider factors extrinsic to the event, even if they modify the event's characteristics—e.g., dust.[11] Nevertheless, this taxonomic clustering can be useful to characterize a population of astrophysical transients with limited information, including subtypes.

Although manual classifications of time-domain phenomena have historically fallen along every level of a proposed taxonomy, automated classifiers have generally considered a fixed level of granularity. This leads to significant preprocessing to construct a labeled data set for training, whereby events not labeled at the leaves are removed and events labeled with even higher granularity are consolidated (e.g., SNe Ib and SNe Ic, typically considered together as SNe Ib/c). This framing of the problem results in an "all-or-nothing" classification task that does not naturally generalize to diverse science goals or observation phases.

ORACLE was designed to leverage every level of a time-domain taxonomy, giving us the ability to derive useful class information even when our model cannot produce confident classifications at the leaves. Adopting a hierarchical approach to classification allows us to "make better mistakes" (L. Bertinetto et al. 2019) in our classifications, and align the granularity of the inference task we consider with the data we have to achieve it. For example, classifying a kilonova (KN) as a rapidly evolving transient before the true nature of the source is known is highly valuable for distinguishing it from more common events and for coordinating follow-up spectroscopy. This coarse classification becomes especially valuable the earlier in an event's evolution it can be made, so as to dedicate more follow-up resources to understand the underlying physics of the object.

### 2.1. Engineering Considerations

Broadly speaking, there are two approaches to achieve hierarchical classification:

1. Build a single classifier that can output the entire classification tree—global hierarchical classification (GHC; L. Bertinetto et al. 2019; J. Schuurmans & F. Frasincar 2023; V. A. Villar et al. 2023).
2. Build a family of classifiers, one for each node in the tree, except for the leaf nodes, which classify between their children using the output of the classifier(s) higher up in the taxonomy—local hierarchical classification (LHC; P. Sánchez-Sáez et al. 2021).

LHCs have multiple drawbacks, including increased training time as a result of using many classifiers and a training process that is sensitive to small changes in the classifier outputs at higher levels in the taxonomy, because of the inherent need to fine-tune several classifiers. A critical engineering concern while developing a production-grade classification pipeline that can operate at the LSST scale is the constraint of running several classifiers in series, as the next classifier in the hierarchy can only run inference after the output from the one higher in the taxonomy is produced. These limitations make LHCs suboptimal for high-throughput, low-latency applications.

By contrast, GHCs have higher throughput and improved inference speed by virtue of having a single classifier.

---

[11] SN interaction with surrounding circumstellar material is an exception, as this extrinsic effect can alter the classification of SNe at every level of the historical SN taxonomy.

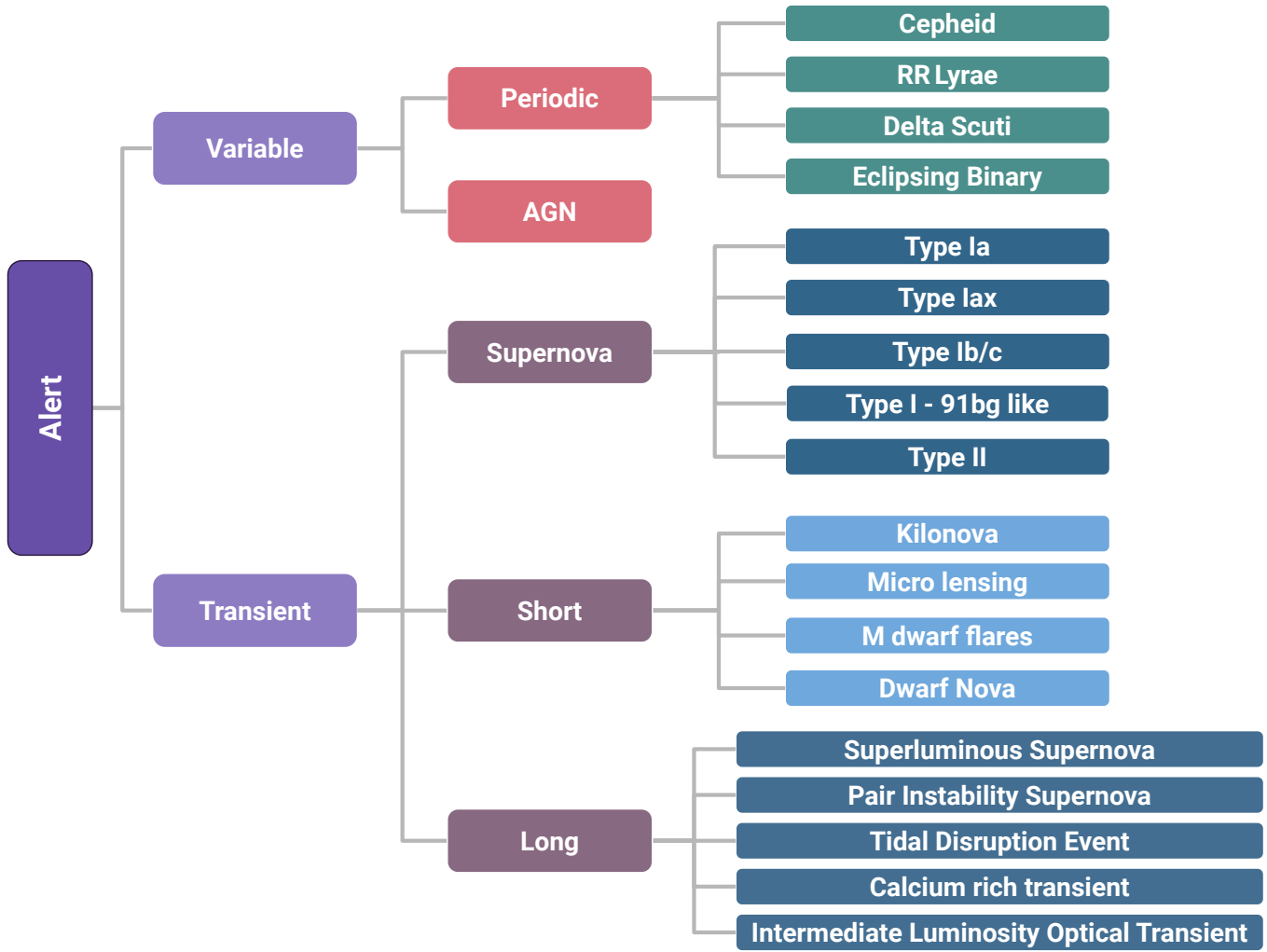

**Figure 1.** Diagram for the taxonomy used in this work. The taxonomy is loosely inspired by the structure of the ELAsTiCC data set, where the leaf nodes are the astrophysical classes, detailed in Table 9.

Additionally, as new classes of astrophysical objects are discovered, the same classifier can be retrained with a new taxonomy and output layer (and no further changes to the architecture of the pipeline), as opposed to adding yet another classifier to the pipeline and further slowing down inference in the case of LHCs.

However, the drawback of the GHC approach is that training requires bespoke loss functions (see Section 6), and while this approach avoids having to tune the performance of several classifiers, specialized hyperparameter tuning may be required to optimize performance across the different layers of a specific taxonomy. For this work, we elected to use a GHC approach, as the upfront increase in development time and complexity was a reasonable trade-off for fast, generalizable performance at inference time.

## 3. Data Set and Taxonomy

We use the simulated "Extended LSST Astronomical Time-series Classification Challenge" (ELAsTiCC;[12] G. Narayan & ELAsTiCC Team 2023) data set for training, validating, and testing ORACLE. Using a simulated data set, such as

ELAsTiCC, avoids the inconsistencies in class labels associated with observed events. ELAsTiCC contains 32 different models, which were mapped to 19 astrophysical classes (documented in Table 9 of Appendix A). This mapping introduces additional diversity to the labeled classes and prevents a classifier from overfitting to the specific details of a transient model.

To classify the ELAsTiCC data set, the natural choice is for the leaf nodes to reflect the true label of the object since any machine learning model developed with it only needs to predict from a known set of classes. The structure of layers higher in the taxonomy is free and can be driven by specific science goals.

There is no universally adopted taxonomy for time-domain astrophysics. Indeed, there is no universal notion of what the "true class" is for many of the most scientifically valuable objects, e.g., objects such as SN 2021foa (D. Farias et al. 2024), which exhibits "flip-flop" behavior, transitioning from IIn-to-Ibn-to-IIn, and such objects might be best identified by anomaly detection techniques (e.g., LAISS and Astro-MCAD; P. D. Aleo et al. 2024; R. Gupta et al. 2025). Additionally, different observers can examine the same observations and come to differing conclusions about the nature of a source.

**Table 1**
Time-independent and Time-dependent Features Used to Train and Test the Neural Network along with Their Transformations

| Time-independent features: | | |
|---|---|---|
| Feature | Description | Representation |
| MWEBV | Milky Way extinction | ⋯ |
| MWEBV_ERR | Error in Milky Way extinction | ⋯ |
| REDSHIFT_HELIO | Best heliocentric redshift. $z$-Spec if available; else $z$-Phot | ⋯ |
| REDSHIFT_HELIO_ERR | Error in best heliocentric redshift | ⋯ |
| HOSTGAL_PHOTOZ | $z$-Phot for the host galaxy if available | ⋯ |
| HOSTGAL_PHOTOZ_ERR | Error in $z$-Phot for the host galaxy | ⋯ |
| HOSTGAL_SPECZ | $z$-Spec for the host galaxy if available | ⋯ |
| HOSTGAL_SPECZ_ERR | Error in $z$-Spec for the host galaxy | ⋯ |
| HOSTGAL_RA | R.A. for the host galaxy | ⋯ |
| HOSTGAL_DEC | decl. for the host galaxy | ⋯ |
| HOSTGAL_SNSEP | Transient-host separation, in arcseconds | ⋯ |
| HOSTGAL_ELLIPTICITY | Ellipticity of the host galaxy | ⋯ |
| HOSTGAL_MAG_[u,g,r,i,z,Y] | $[u, g, r, i, z, Y]$—band magnitudes for the host galaxy | ⋯ |
| MW_plane_flag | Flag to mark location in the Milky Way plane | 1 if in-plane, else 0 |
| ELAIS_S1_flag | Flag to mark location in the ELAIS S1 field | 1 if in-field, else 0 |
| XMM-LSS_flag | Flag to mark location in the XMM-LSS field | 1 if in-field, else 0 |
| Extended_Chandra_Deep_Field_South_flag | Flag to mark the location in the Extended Chandra Deep Field | 1 if in-field, else 0 |
| COSMOS_flag | Flag to mark location in the COSMOS field | 1 if in-field, else 0 |

| Time-dependent features: | | |
|---|---|---|
| Feature | Description | Transformation |
| FLUXCAL | Calibrated flux value from SNANA | Remove saturation(s), then FLUXCAL/1000 |
| FLUXCAL_ERR | Uncertainty on FLUXCAL from SNANA | Remove saturation(s), then FLUXCAL_ERR/1000 |
| MJD | Modified Julian date | [MJD − min(MJD)]/100 |
| BAND | Passband name | Mean wavelength of the passband in $\mu$m (see Table 2) |
| PHOTFLAG | Flags for detections and nondetections | 1 for detection, 0 for nondetections. Remove saturations |

**Note.** Redshift distributions are shown in Figure 2.

A consideration while designing the taxonomy is which basis to adopt for classification at every level. When defining higher levels, we have aimed for consistency: for instance, Galactic versus extragalactic classification was not chosen at the top level, as it would involve mixing a spatial basis for classification with a temporal one at the leaves (periodic versus stochastic variability). Another consideration is ensuring that categories are distinct at every level: some objects, such as Cepheids, can have either a galactic or extragalactic origin, resulting in an artificial inflation in the number of leaf nodes without added information. The final taxonomy adopted in this work (Figure 1) is observationally motivated and similar to that used by LSST alert brokers, such as ALeRCE (P. Sánchez-Sáez et al. 2021).

We stress that, regardless of the details of the taxonomy, the hierarchical classification approach implemented in this work has broad applicability to other use cases and can serve as a proof of concept for other studies, e.g., morphological classification of galaxies. Since classification based on observational characteristics often informs action, the taxonomy is continually evolving, as new observations naturally revise the taxonomy—e.g., type Ia supernovae are now subdivided into "Ia-CSM," "Iax," and other categories (R. J. Foley et al. 2013; J. M. Silverman et al. 2013).

As the adopted taxonomy is observationally driven, the latent space used to encode the information from astrophysical sources must capture the observational differences between object classes. We combine a mixture of time-dependent and time-independent features, described in Section 4, to characterize these events. While most of the features we define in Table 1 are well defined in the literature, we note that the FLUXCAL feature is specific to the SNANA (R. Kessler et al. 2009) software package used to simulate the ELAsTiCC data set and is defined as follows:

$$\text{mag} = 27.5 - 2.5 \cdot \log_{10}(\text{FLUXCAL}).  \tag{1}$$

The 5 time-dependent features and 23 time-independent features used for training the model are summarized in Tables 1 and 2. Most of these features are used either as provided in ELAsTiCC or with minor transformations. We also introduce a new set of flags to indicate if the coordinates of the source fall within one of the LSST Deep Drilling Fields (DDFs). Data from the DDFs will differ from wide–fast–deep (WFD) data both in the distribution of transient classes detected and in the cadence of the photometry. They will also have a different distribution of redshifts (see Figure 2). By adding indicator variables, we encourage the model to distinguish between these surveys and adjust its weights accordingly.

**Table 2**
Mean Wavelengths for the Six LSST Passbands

| Band | Mean Wavelength |
| --- | --- |
| LSST $u$ | 0.360 $\mu$m |
| LSST $g$ | 0.476 $\mu$m |
| LSST $r$ | 0.622 $\mu$m |
| LSST $i$ | 0.755 $\mu$m |
| LSST $z$ | 0.870 $\mu$m |
| LSST $Y$ | 1.015 $\mu$m |

Intuitively, taxonomies that distinguish based on observable characteristic(s) of the light curve and host galaxy, such as the one used here, should perform better than ones that distinguish strictly based on the underlying physics of the object. This is likely because distinct physical processes do not always imprint unique signatures on broadband photometry. A detailed study of how classification performance is affected by the choice of taxonomy is beyond the scope of this work.

## 4. Data Augmentation

First, the data set was divided into training and testing sets using a 70:30% split. Then, we keep at most 40,000 samples of a class in the training set and at most 20,000 samples of a class in the testing set, where a sample is defined as a tuple of a light curve array and a static features array. We introduce the per-class sample limit to achieve a balance between model performance and training/inference time. From the modified training set, an additional 5% of the data was reserved for validation. The number of samples from each class for the training, validation, and testing sets is documented in Table 3.

Since the objective of this work is to build a real-time system capable of classifying partial-phase light curves, we explicitly train ORACLE on augmented data. For each light curve in the training set, we sample a fraction, $f$, from $U$ (0.1, 1). Then, we truncate the light curve to this fraction of the original number of observations, where the number of observations includes both detections and nondetections. This data augmentation step is repeated at the start of each training epoch to ensure that the model is trained on light curves at several stages of an event's evolution, and to make the model robust to incomplete photometry.

For the validation set, we augmented each light curve to 10%, 40%, 60%, and 100% of its original number of observations. This results in a new augmented validation set with 4 times as many samples and is representative of events at different phases of their evolution.

For the testing set, each light curve was augmented to observations within $d$ days since first detection (or trigger), where $d = 2^n$, $\forall\, n \in \mathbb{Z} \cap [0, 10]$.

The difference in data augmentation method between the training and testing set is to enable the evaluation of the model's performance at earlier phases than considered in training, a meaningful strategy to demonstrate performance on out-of-distribution events. Finally, we padded all the time series to a length of 500 in order to ensure that every light curve could be ingested into the classifier in batches.

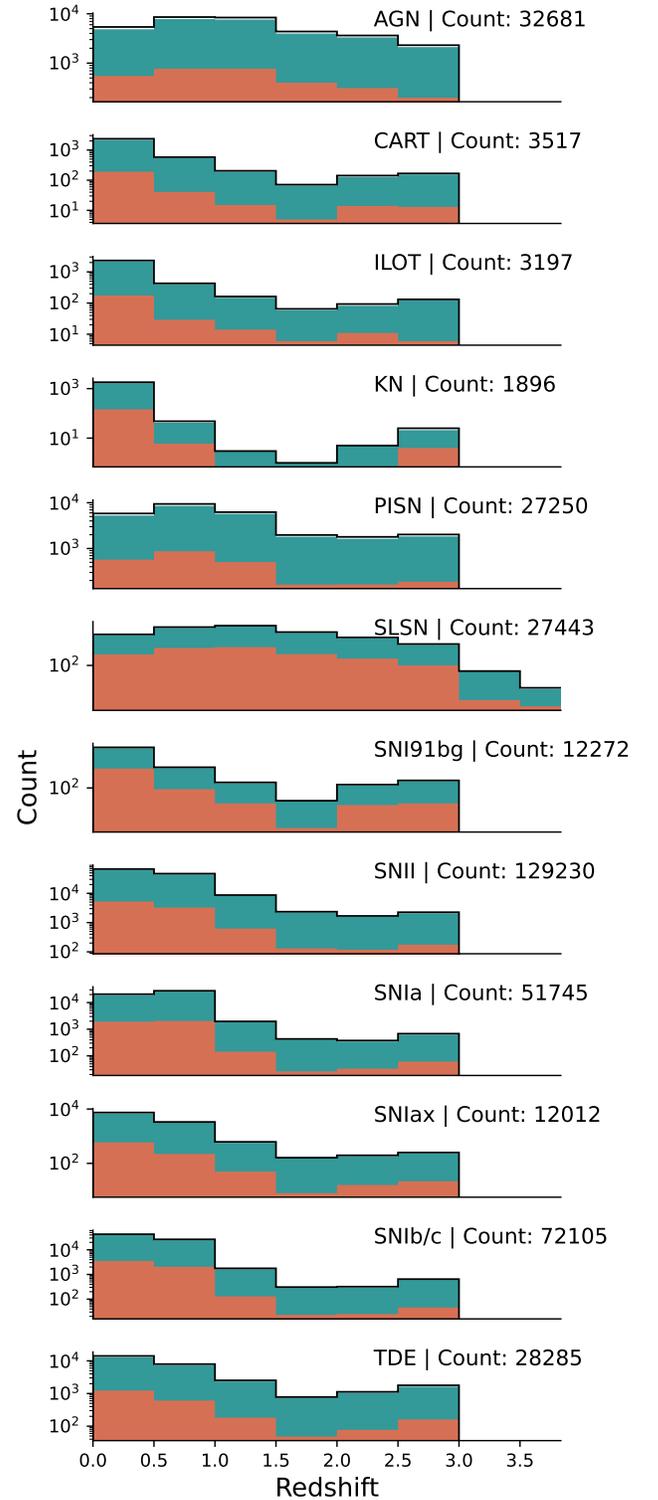

**Figure 2.** Histograms of redshifts (spectroscopic if available, else photometric) for extragalactic classes in the test set used for this work. Histograms for events in the LSST DDFs (shown in orange) indicate a higher fraction of sources at higher redshifts compared to WFD (shown in teal). The combined distribution for both WFD and DDF is shown with a solid black line. The total number of sources in each class is labeled.

This brings the total number of training samples to $\sim$450,000 ($\times$1 after phase augmentation), the total number of validation samples to $\sim$24,000 ($\times$4 after phase augmentation), and the total number of testing samples to $\sim$225,000 ($\times$11 after phase augmentation).

| Class | Train | Test | Validation |
|---|---|---|---|
| AGN | 38,113 | 20,000 | 1887 |
| CaRT | 7807 | 3517 | 400 |
| Cepheid | 13,088 | 5901 | 683 |
| Delta Scuti | 19,611 | 8849 | 1039 |
| Dwarf Novae | 7608 | 3439 | 417 |
| EB | 38,036 | 20,000 | 1964 |
| ILOT | 7090 | 3197 | 371 |
| KN | 4211 | 1896 | 215 |
| M dwarf flare | 1780 | 796 | 79 |
| PISN | 37,996 | 20,000 | 2004 |
| RR Lyrae | 13,278 | 6014 | 755 |
| SLSN | 37,993 | 20,000 | 2007 |
| SNI91bg | 27,207 | 12,272 | 1430 |
| SNII | 38,018 | 20,000 | 1982 |
| SNIa | 38,000 | 20,000 | 2000 |
| SNIax | 26,610 | 12,012 | 1420 |
| SNIb/c | 37,935 | 20,000 | 2065 |
| TDE | 38,023 | 20,000 | 1977 |
| uLens | 16,652 | 7537 | 940 |
| Total | 449,056 | 225,430 | 23,635 |

We trained the final model on the Illinois Campus Cluster using an NVIDIA H100 GPU. Each training epoch took ∼40 s, bringing the total training time to ∼12 hr. Furthermore, all the analyses reported in this work took ∼4 hr on the same hardware, bringing the total time for the training and evaluation of the model to ∼16 hr. Due to the resource-intensive process of model training, data augmentation, and evaluation described above, we resort to using frozen training, validation, and testing sets instead of using cross validation.

## 5. Model Architecture

There are several approaches to classifying irregularly sampled light curves. Some popular approaches involve:

1. Extracting statistical features from the light curve and using them for classification with machine learning models such as random forests (P. Sánchez-Sáez et al. 2021) or multilayer perceptrons (V. A. Villar et al. 2023).
2. Directly use the time-series photometry for classification, using some type of RNN (D. Muthukrishna et al. 2019).
3. Computing an encoded representation of the light curves (e.g., with variational autoencoders (VAE), like VRAENN; V. A. Villar et al. 2021), which can then be used with a downstream classifier. In modern transformer-based models, a positional encoding is used to preserve phase information (G. Cabrera-Vives et al. 2024).
4. Interpolating light curves and then using the interpolated data for classification (K. Boone 2019).

In this work, the time-series data were processed with an RNN, after minor transformations (see Table 1). While discussing the merits of each approach in detail is outside the scope of this paper, it is worth noting that there is much discussion about the interpretability of features extracted by neural networks during the training process. In particular, neural networks are known to extract features that are often nebulous and difficult to explain, especially in the context of physical systems (A. Shahroudnejad 2021). While machine learning models such as random forest and support vector machine are generally more explainable because their decision-making processes can be traced back to simple rules, they are routinely outperformed by more sophisticated deep-learning methods in the context of astrophysical classification (H. Klimczak et al. 2022; G. Cabrera-Vives et al. 2024).

Interpretability aside, RNNs have been shown to accurately classify time-series data. In particular, ORACLE (see Figure 3) is inspired by the structure of the RAPID classifier (D. Muthukrishna et al. 2019), since it has been demonstrated to work reliably with astrophysical data. In order to deal with light curves of varying length, we first pad our time-series data (with zeros) to ensure that we can train in batches. In the model architecture, we use a masking layer to ignore the padded values during training and inference. We also use a flag value (of −9) to indicate missing static features, which are also masked off in the second input branch for the model.

In addition to the full model, we also trained a model that only uses time-dependent features, called ORACLE-lite. This model does not include the static feature branch shown in Figure 3.

Given how we have structured our code base, it would be trivial to retrain future versions of ORACLE with more sophisticated architectures.

## 6. Loss Function

The model must learn to make predictions along a specified classification hierarchy, and the selected loss function must facilitate this goal. For this work, we use a modified version of the hierarchical cross-entropy (HXE) loss function (L. Bertinetto et al. 2019), known as the weighted HXE (WHXE; V. A. Villar et al. 2023), which introduces an additional weighing factor with the goal of improving the model's performance on unbalanced data sets, which are typically found in astrophysics. We summarize the HXE and the WHXE below.

First, the probability of class $C$ in the hierarchy is described as

$$p(C) = \prod_{l=0}^{h(C)-1} p(C^{(l)}|C^{(l+1)}), \qquad (2)$$

where $h(C)$ is the height of node $C$ in the hierarchy. This means that $C^{(0)}$ is a leaf node, while $C^{(H)}$ is the root node, assuming $H$ is the height of the tree. Since all classifications begin with an Alert, we use $p(C^{(H)}) = p(\text{Alert}) = 1$. We can also express the conditional probabilities as

$$p(C^{(l)}|C^{(l+1)}) = \frac{\sum_{A \in \text{Leaves}(C^{(l)})} p(A)}{\sum_{B \in \text{Leaves}(C^{(l+1)})} p(B)}, \qquad (3)$$

where Leaves($C$) represents the set of leaves for the sub tree starting at Node $C$. Next, an additional term was introduced to weigh the losses at different nodes, based on where they appear in the hierarchy

$$\lambda(C^{(l)}) = \exp(-\alpha d(C)). \qquad (4)$$

Here, $d(C)$ represents the depth of the node $C$ in the hierarchy, and $\alpha$ is a free parameter that can be tweaked to affect the depth whose performance gets prioritized during training. For example, $\alpha = 0$ weighs all nodes equally, while $\alpha > 0$ preferentially weighs nodes higher in the taxonomy.

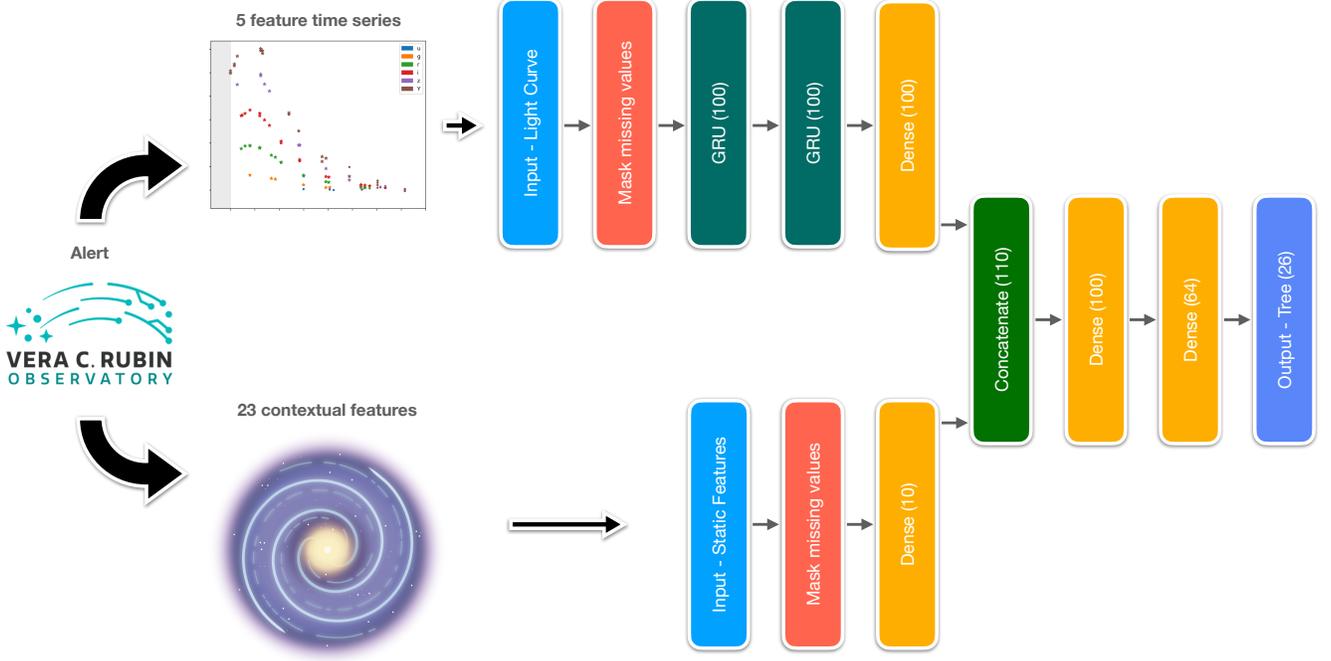

**Figure 3.** Architecture of the RNN model used for this work.

Incorporating these elements, L. Bertinetto et al. (2019) defined the HXE loss as

$$\text{Ł}_{\text{HXE}}(p, C) = -\sum_{l=0}^{h-1} \lambda(C^{(l)}) \log p(C^{(l)}|C^{(l+1)}), \quad (5)$$

where $C$ is the node of the true class. Since astrophysical data sets are often class-imbalanced, V. A. Villar et al. (2023) modified the HXE loss by adding a factor to weight the losses based on the number of occurrences of each class in the data set. The new weight term is defined as

$$W(C^{(l)}) = \frac{N}{N_{\text{nodes}} \cdot N_{\text{c}}} \quad (6)$$

where $N$ is the total number of events in the data set, $N_{\text{nodes}}$ is the number of unique classes, and $N_{\text{c}}$ is the number of events of class $C$. Combining these elements, V. A. Villar et al. (2023) formulated the WHXE loss function as follows:

$$\text{Ł}_{\text{WHXE}}(p, C) = -\sum_{l=0}^{h-1} W(C^{(l)}) \lambda(C^{(l)}) \log p(C^{(l)}|C^{(l+1)}). \quad (7)$$

We have included a generic implementation for both the HXE and WHXE loss functions for TensorFlow/Keras in the repository for this project.[13]

## 7. Training

While we conducted some experiments to find a set of hyperparameter values that worked well for training ORACLE, no formal hyperparameter tuning was used to achieve the results in Section 8. The hyperparameter values used for training the model are documented in Table 4.

**Table 4**
Hyperparameters Used for Training ORACLE

| Parameter | Value |
| --- | --- |
| Epochs | 1000 |
| Batch Size | 1024 |
| Learning rate | 2e-4 |
| Nonlinearity | tanh, ReLU |
| Optimizer | Adam |
| Alpha (for loss function) | 0.5 |

We recorded the mean loss on the augmented validation set for each epoch of training. Then, the model with the lowest mean loss was used (Figure 4).

We use a learning rate with an exponential decay schedule with a decay rate of 0.9 for every 10,000 steps, which mimics annealing, and allows the network to avoid settling in local minima early in the training process.

As mentioned in Section 5, we train another version of the model, ORACLE-lite, using only the time-dependent features. This eliminates the dependence of the classifier on features like redshift, extinction, and host galaxy properties.

## 8. Results

Given the hierarchical nature of the problem, we can separately consider the performance of the network at every layer in our defined taxonomy. We use the label level_1 to refer to the classification performance at a depth of 1 in the taxonomy (the highest level of classification), level_2 to indicate performance at a depth of 2, and leaf to refer to classification at the leaves of the taxonomy (the most granular level of classification). We note here that the active galactic nucleus (AGN) class appears at both the level 2 and leaf classification since it is both a leaf class and has a depth of 2 in the taxonomy (Figure 3).

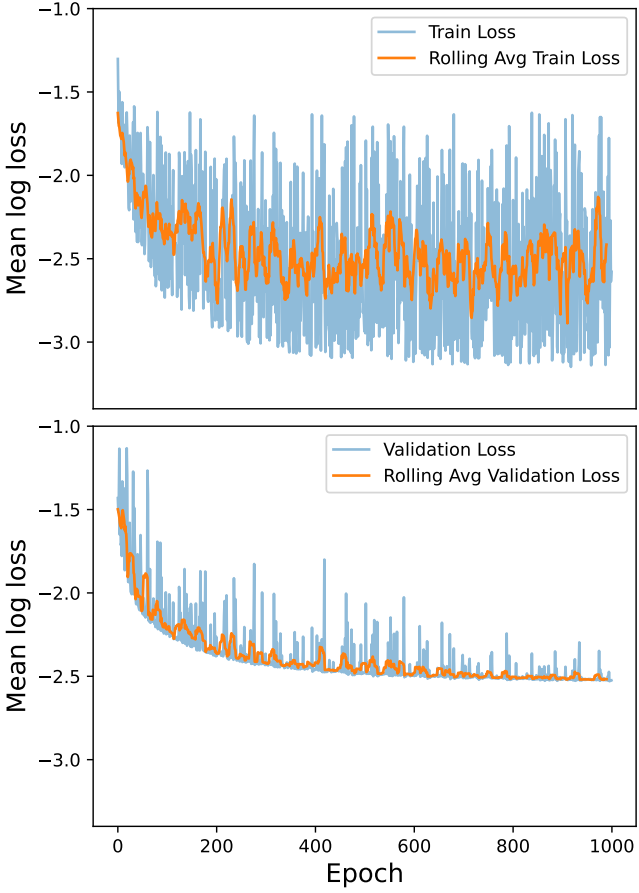

**Figure 4.** Mean train (top) and validation (bottom) losses for every epoch in the model's training history. A window size of 10 was used for computing the rolling mean. The large fluctuations in the training loss (and to a lesser extent, the validation loss) are likely due to large variations in the training set, as a result of the augmentations made at the start of every epoch.

We start this discussion with the full version of ORACLE which has been trained on both time-dependent and time-independent features. We report the performance (precision, recall, and F-1 score, defined in the equations below) as a function of the days (after trigger) in Figures 5 and 6. We also tabulate the efficacy of the model at different phases of the light curves' evolution in Table 5. The definition of precision, recall, and F-1 score is as follows:

$$\text{Precision} = \frac{\text{True Positive}}{\text{True Positive} + \text{False Positive}}, \quad (8)$$

$$\text{Recall} = \frac{\text{True Positive}}{\text{True Positive} + \text{False Negative}}, \quad (9)$$

$$\text{F-1 score} = 2 \cdot \frac{\text{Precision} \cdot \text{Recall}}{\text{Precision} + \text{Recall}}. \quad (10)$$

While reporting the results, we make use of both macro and weighted (or micro) averaged metrics. A macro average gives equal weight to all classes, regardless of how many sources from the class were tested. By comparison, the micro/weighted average weighs the metric for each class by the number of sources belonging to that class. Microaverages are biased toward classes with more objects; while these metrics provide a useful high-level view of model performance, they provide little insight into the model's performance on rarer classes (and are highly sensitive to the class distribution of the training/validation set, which may not be representative). For this reason, we choose to focus on macroaveraged metrics.

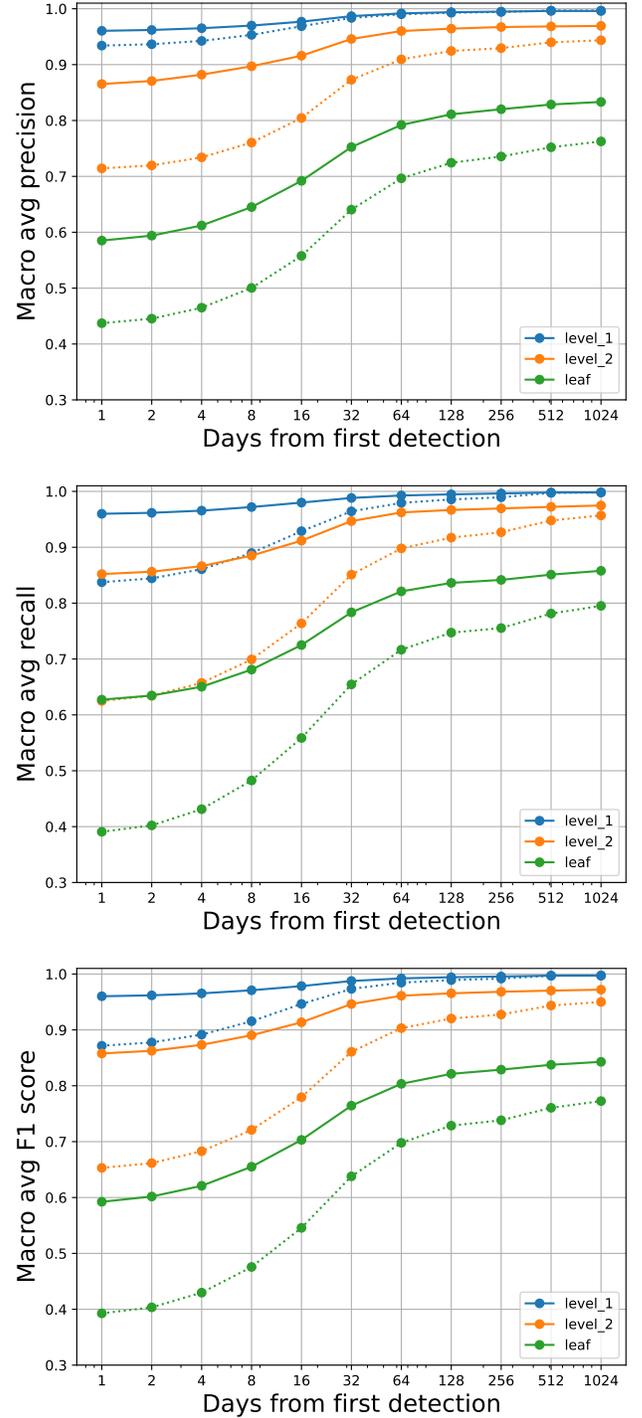

**Figure 5.** Macroaveraged precision (top), recall (middle), and F-1 score (bottom) as a function of the number of days since event trigger. Performance for the full model (trained using both time-dependent and time-independent features) is plotted using solid lines. Performance for ORACLE-lite (trained using only time-dependent features) is plotted using dotted lines.

The earliest phase at which we evaluated performance was 1 day after the event trigger. At this time, most sources had fewer than four total detections. A small fraction of sources (<2%) had >4 detections within this period, either because they are in one of the LSST DDFs or due to serendipitous overlap of tilings. This means that we can provide confident

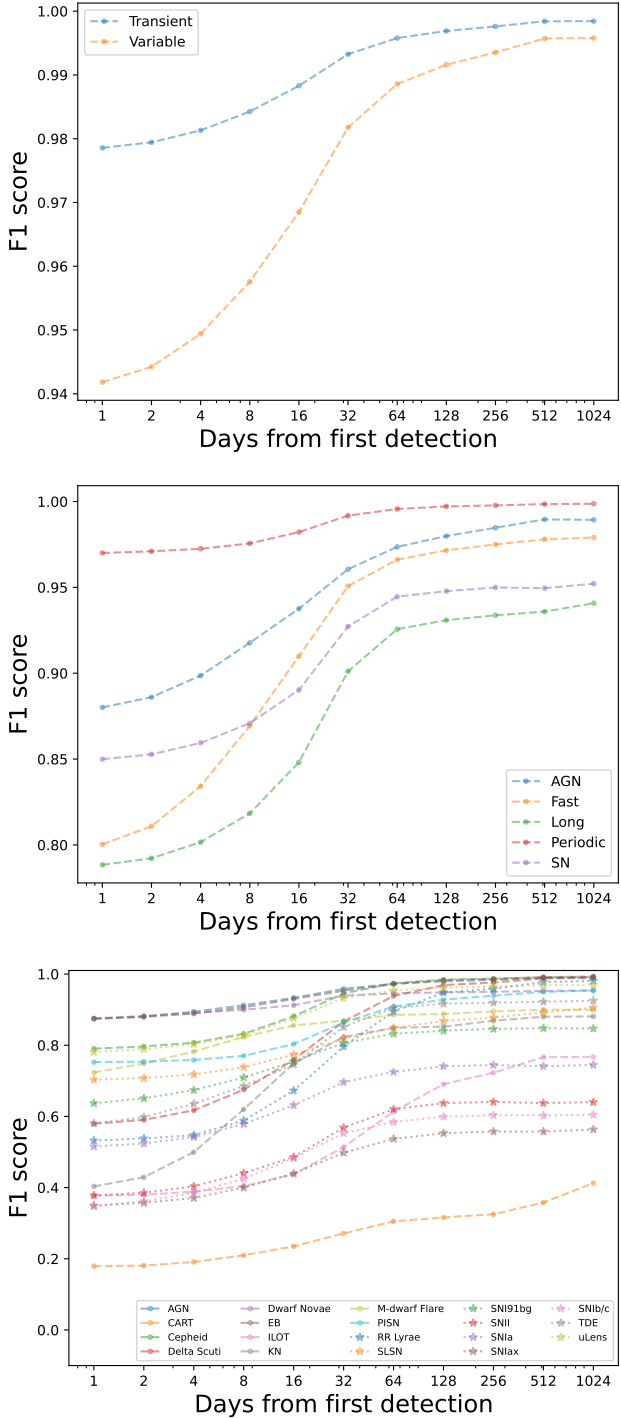

**Figure 6.** Per-class F-1 scores for the level_1 (top), level_2 (middle), and the leaf (bottom) as a function of the number of days since event trigger.

level_1 and level_2 classifications, with >85% precision (Figure 5) and recall, within 24 hr of the first detection. Example light curves that have been correctly classified, along with the true class probabilities' evolution over time, have been shown in Figures 19 and 20.

Confusion matrices at Trigger + 2, 8, 64, and 1024 days, at the three levels described above, are presented in Figures 7–9, respectively, while the corresponding receiver operating characteristic (ROC) curves are presented in Figures 14–16, respectively. Trigger + 2, 8, and 64 days were chosen to facilitate comparisons with other models in the literature, while Trigger + 1024 days was chosen to demonstrate ORACLE's performance when "complete" light curves are available.

Finally, Figures 17 and 18 show the evolution of the mean class scores with phase, organized by the true class.

We report level 1, 2, and leaf macro F-1 scores of 0.96, 0.86, and 0.60, respectively, at Trigger + 2 days. These figures improve to 0.97, 0.89, and 0.66, at Trigger + 8 days, before finally reaching ∼ 1, 0.97, and 0.84 at Trigger + 1024 days.

We ran additional tests to analyze the pretrigger performance of the models. For this exercise, we ran inference using two sets of data. For the first set, all the static features were removed, leaving only the pretrigger nondetections for classification. For the second set, all the pretrigger nondetections were removed, leaving just the static features. As anticipated, in either case, the performance degraded severely when compared with the baseline (using both static features and pretrigger nondetections), likely because the model was never explicitly trained on such data. Table 6 tabulates the pretrigger performance metrics for ORACLE. We note that it is, in principle, possible to train models that can make classifications exclusively based on contextual information alone (A. Gagliano et al. 2021) when they are specifically trained to perform these tasks.

As mentioned in Sections 5 and 7, we train an additional model, ORACLE-lite, which only uses time-dependent features for training and classification. This model achieves level 1, 2, and leaf macro F-1 scores of ∼1, 0.94, 0.77 at Trigger + 1024 days. While we do not provide detailed statistics for the performance of this model, Figures 5 and 10 show comparisons between the two models we train for this paper. While the light curve-only approach is particularly important for smaller surveys where it may not be feasible to compute features such as host galaxy matches and photometric redshift estimates, it also highlights the importance of these static features in improving the effectiveness of the models, especially at early times when limited photometry is available.

Importantly, ORACLE achieves classification performance comparable to other state-of-the-art classifiers, some of which have considerably more sophisticated transformer-based architectures (G. Cabrera-Vives et al. 2024), all while being more versatile by virtue of being able to make predictions at every level in the taxonomy (e.g., BHRF P. Sánchez-Sáez et al. 2021). Table 8 compares the relative performance of ORACLE with the state-of-the-art models from the literature.

This blend of classification performance, a hierarchical structure, and real-time capabilities makes ORACLE a unique addition to the ecosystem of classification tools built in preparation for LSST.

### 8.1. Inference Runtime Performance

Much like the classification performance metrics comparisons shown in Table 8, it is difficult to do an apples-to-apples comparison for the inference runtime performance without running comprehensive tests on multiple classifiers on the same hardware. Here, we adhere to the statistics reported by G. Cabrera-Vives et al. (2024) for ATAT, albeit with different hardware, in the hopes of standardizing the metrics reported. We run inference on 20,000 samples using a batch size of 1 and 2000, and record the average time required for each sample (Table 7). We used an NVIDIA H100 to run the GPU

**Table 5**
Performance Summary with Precision ($p$), Recall ($r$), and F-1 Scores ($f1$) for the Test Set

| Class | $p_{2d}$ | $r_{2d}$ | $f1_{2d}$ | $p_{8d}$ | $r_{8d}$ | $f1_{8d}$ | $p_{64}$ | $r_{64d}$ | $f1_{64d}$ | $p_{1024d}$ | $r_{1024d}$ | $f1_{1024d}$ |
|---|---|---|---|---|---|---|---|---|---|---|---|---|
| **Performance at Level 1:** | | | | | | | | | | | | |
| Transient | 0.98 | 0.98 | 0.98 | 0.99 | 0.98 | 0.98 | 1.00 | 1.00 | 1.00 | 1.00 | 1.00 | 1.00 |
| Variable | 0.94 | 0.94 | 0.94 | 0.95 | 0.96 | 0.96 | 0.99 | 0.99 | 0.99 | 0.99 | 1.00 | 1.00 |
| accuracy | ... | ... | 0.97 | ... | ... | 0.98 | ... | ... | 0.99 | ... | ... | 1.00 |
| macro avg | 0.96 | 0.96 | 0.96 | 0.97 | 0.97 | 0.97 | 0.99 | 0.99 | 0.99 | 1.00 | 1.00 | 1.00 |
| weighted avg | 0.97 | 0.97 | 0.97 | 0.98 | 0.98 | 0.98 | 0.99 | 0.99 | 0.99 | 1.00 | 1.00 | 1.00 |
| **Performance at Level 2:** | | | | | | | | | | | | |
| AGN | 0.93 | 0.85 | 0.89 | 0.94 | 0.90 | 0.92 | 0.97 | 0.98 | 0.97 | 0.98 | 1.00 | 0.99 |
| Fast | 0.83 | 0.80 | 0.81 | 0.89 | 0.85 | 0.87 | 0.96 | 0.97 | 0.97 | 0.97 | 0.99 | 0.98 |
| Long | 0.83 | 0.76 | 0.79 | 0.85 | 0.79 | 0.82 | 0.94 | 0.91 | 0.93 | 0.94 | 0.95 | 0.94 |
| Periodic | 0.95 | 0.99 | 0.97 | 0.96 | 0.99 | 0.98 | 0.99 | 1.00 | 1.00 | 1.00 | 1.00 | 1.00 |
| SN | 0.83 | 0.88 | 0.85 | 0.85 | 0.90 | 0.87 | 0.94 | 0.95 | 0.94 | 0.96 | 0.94 | 0.95 |
| accuracy | ... | ... | 0.86 | ... | ... | 0.88 | ... | ... | 0.95 | ... | ... | 0.96 |
| macro avg | 0.87 | 0.86 | 0.86 | 0.90 | 0.89 | 0.89 | 0.96 | 0.96 | 0.96 | 0.97 | 0.97 | 0.97 |
| weighted avg | 0.86 | 0.86 | 0.86 | 0.88 | 0.88 | 0.88 | 0.95 | 0.95 | 0.95 | 0.96 | 0.96 | 0.96 |
| **Performance at the Leaves:** | | | | | | | | | | | | |
| AGN | 0.90 | 0.86 | 0.88 | 0.92 | 0.91 | 0.91 | 0.97 | 0.98 | 0.97 | 0.98 | 1.00 | 0.99 |
| CaRT | 0.14 | 0.24 | 0.18 | 0.17 | 0.27 | 0.21 | 0.30 | 0.31 | 0.31 | 0.37 | 0.46 | 0.41 |
| Cepheid | 0.72 | 0.90 | 0.80 | 0.76 | 0.92 | 0.83 | 0.96 | 0.99 | 0.97 | 0.99 | 1.00 | 0.99 |
| Delta Scuti | 0.64 | 0.55 | 0.59 | 0.69 | 0.66 | 0.68 | 0.93 | 0.95 | 0.94 | 0.99 | 0.99 | 0.99 |
| Dwarf Novae | 0.90 | 0.86 | 0.88 | 0.91 | 0.89 | 0.90 | 0.94 | 0.95 | 0.95 | 0.94 | 0.97 | 0.95 |
| EB | 0.91 | 0.85 | 0.88 | 0.94 | 0.88 | 0.91 | 0.98 | 0.96 | 0.97 | 0.99 | 0.99 | 0.99 |
| ILOT | 0.32 | 0.47 | 0.38 | 0.35 | 0.49 | 0.40 | 0.52 | 0.75 | 0.61 | 0.68 | 0.87 | 0.77 |
| KN | 0.30 | 0.75 | 0.43 | 0.49 | 0.85 | 0.62 | 0.76 | 0.97 | 0.85 | 0.82 | 0.96 | 0.88 |
| M dwarf flare | 0.67 | 0.86 | 0.75 | 0.77 | 0.88 | 0.82 | 0.85 | 0.93 | 0.89 | 0.88 | 0.92 | 0.90 |
| PISN | 0.75 | 0.76 | 0.75 | 0.79 | 0.75 | 0.77 | 0.90 | 0.92 | 0.91 | 0.95 | 0.96 | 0.95 |
| RR Lyrae | 0.45 | 0.66 | 0.54 | 0.52 | 0.68 | 0.59 | 0.87 | 0.92 | 0.90 | 0.98 | 0.98 | 0.98 |
| SLSN | 0.76 | 0.66 | 0.71 | 0.80 | 0.68 | 0.74 | 0.91 | 0.80 | 0.85 | 0.91 | 0.90 | 0.91 |
| SNI91bg | 0.55 | 0.79 | 0.65 | 0.62 | 0.83 | 0.71 | 0.78 | 0.89 | 0.83 | 0.80 | 0.91 | 0.85 |
| SNII | 0.45 | 0.34 | 0.39 | 0.52 | 0.38 | 0.44 | 0.70 | 0.56 | 0.62 | 0.75 | 0.56 | 0.64 |
| SNIa | 0.48 | 0.57 | 0.52 | 0.53 | 0.64 | 0.58 | 0.69 | 0.76 | 0.72 | 0.74 | 0.75 | 0.74 |
| SNIax | 0.33 | 0.39 | 0.36 | 0.37 | 0.44 | 0.40 | 0.49 | 0.59 | 0.54 | 0.50 | 0.65 | 0.56 |
| SNIb/c | 0.42 | 0.32 | 0.36 | 0.47 | 0.39 | 0.42 | 0.61 | 0.56 | 0.58 | 0.65 | 0.57 | 0.60 |
| TDE | 0.69 | 0.53 | 0.60 | 0.75 | 0.63 | 0.69 | 0.92 | 0.89 | 0.91 | 0.92 | 0.93 | 0.93 |
| uLens | 0.89 | 0.71 | 0.79 | 0.90 | 0.77 | 0.83 | 0.96 | 0.94 | 0.95 | 0.98 | 0.96 | 0.97 |
| accuracy | ... | ... | 0.62 | ... | ... | 0.66 | ... | ... | 0.81 | ... | ... | 0.84 |
| macro avg | 0.59 | 0.63 | 0.60 | 0.65 | 0.68 | 0.66 | 0.79 | 0.82 | 0.80 | 0.83 | 0.86 | 0.84 |
| weighted avg | 0.64 | 0.62 | 0.62 | 0.68 | 0.66 | 0.67 | 0.81 | 0.81 | 0.81 | 0.85 | 0.84 | 0.84 |

**Note.** All values were computed after augmenting the light curves to Trigger + 2, 8, 64, and 1024 days (represented using the subscript). Trigger + 2, 8, and 64 days were chosen to facilitate comparisons with other models in the literature, while Trigger + 1024 days was chosen to demonstrate ORACLE's performance when "complete" light curves are available.

timing test and an Apple M1 Pro (using a single core) to run the CPU timing test.

From Table 7, we can infer that, with a batch size of 2000, we can classify each sample in about $1.8 \times 10^{-3}$ s using a CPU, which translates to a rate of ~555 alerts/second/core. This places ORACLE comfortably above the ~100 alerts/second/core "LSST real-time ready" threshold suggested in T. Allam et al. (2023), and scalable to expected LSST throughputs.

Since we did not conduct formal hyperparameter searches, we have not attempted to optimize the model architecture for runtime performance. It is possible that a more lightweight model with fewer trainable parameters can deliver similar classification accuracy.

### 8.2. Model Comparisons

In this section, we qualitatively and quantitatively compare ORACLE with other photometric classifiers from the literature. It is crucial to note that the goal of this section is not to provide a detailed comparison of the differences in performance of these classifiers since there are far too many

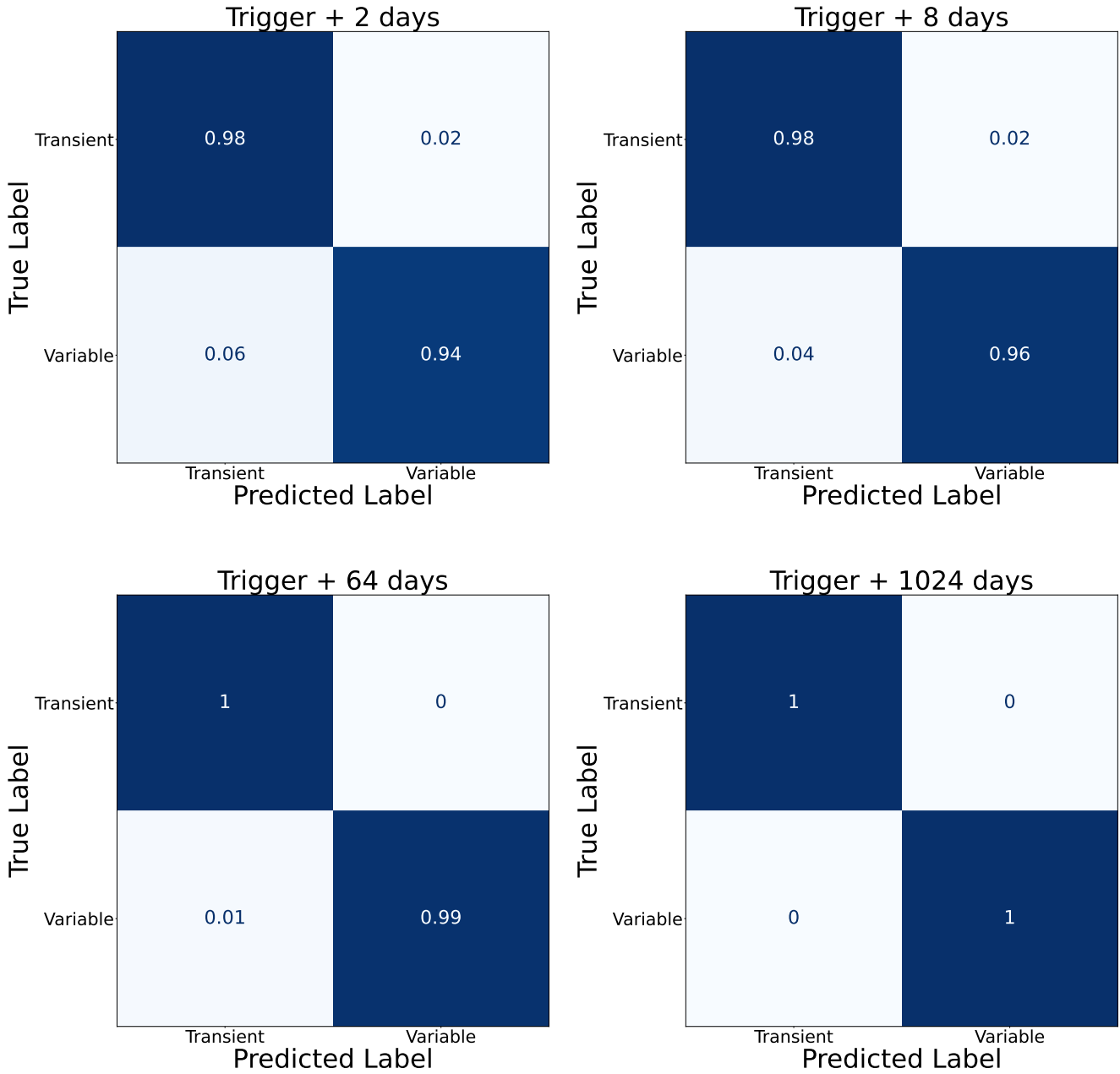

**Figure 7.** Confusion matrices for classification at level 1, as a function of the number of days since event trigger.

Table 6
F-1 Scores before First Detection: (1) Only Nondetections, (2) Only Static Features, (3) Both Static Features and Nondetections

| Features | Level 1 F-1 | Level 2 F-1 | Leaf F-1 |
|----------|-------------|-------------|----------|
| ND only | 0.79 | 0.21 | 0.06 |
| Static only | 0.50 | 0.26 | 0.05 |
| Both | 0.94 | 0.83 | 0.54 |

Table 7
Average Runtime Comparison for Inference between ORACLE (Using an NVIDIA H100 GPU/Apple M1 Pro CPU) and ATAT (Using an NVIDIA A100 GPU/AMD EPYC 7662)

| | ORACLE | ATAT |
|---|---|---|
| **GPU** | | |
| 2000 light curves per batch | $1.3 \times 10^{-4}$ s | $4.75 \times 10^{-4}$ s |
| 1 light curve per batch | $1.0 \times 10^{-2}$ s | $8.14 \times 10^{-3}$ s |
| **CPU** | | |
| 2000 light curves per batch | $1.8 \times 10^{-3}$ s | $6.4 \times 10^{-3}$ s |
| 1 light curve per batch | $1.9 \times 10^{-2}$ s | ... |

**Note.** ORACLE: using an NVIDIA H100 GPU/Apple M1 Pro CPU; ATAT: using an NVIDIA A100 GPU/AMD EPYC 7662. Differences in performance are, at least in part, due to differences in the hardware used for inference.

variables (such as the number of classes, data sets, and granularity) to control, given our high-level treatment. Instead, we focus on providing a broad overview and placing ORACLE in the context of the current landscape of photometric classifiers.

In Table 8, we report the classification performance of ORACLE along with other popular real-time classifiers: ATAT

**Table 8**
Performance Comparison between ORACLE (this Work), ATAT, RAPID, and SCONE

| Statistic | ORACLE[a] | ATAT[b] | RAPID[c] | SCONE[d] | SUPERNNOVA[e] |
|---|---|---|---|---|---|
| Trigger + 2 days, Bottom-level macro F-1 | 0.60 | ≈0.65 | ⋯ | ⋯ | ⋯ |
| Trigger + 2 days, Mid-level ROC macro-AUC | 0.97 | ⋯ | ⋯ | ≈0.93 | ⋯ |
| Trigger + 2 days, Bottom-level ROC macro-AUC | 0.96 | ⋯ | 0.93 | ⋯ | ⋯ |
| Trigger + 32 days, Mid-level ROC macro-AUC | 0.99 | ⋯ | ⋯ | ≈0.98 | ⋯ |
| Trigger + 32 days, Bottom-level ROC macro-AUC | 0.98 | ⋯ | ≲0.97 | ⋯ | ⋯ |
| Trigger + 1024 days, Mid-level macro F-1 | 0.97 | ⋯ | ⋯ | ⋯ | 0.95 |
| Trigger + 1024 days, Bottom-level macro F-1 | 0.84 | 83.5 ± 0.6 | ⋯ | ⋯ | ⋯ |
| Trigger + 1024 days, Mid-level ROC macro-AUC | >0.99 | ⋯ | ⋯ | >0.99 | ⋯ |

**Notes.** Note that, due to the flat nature of ATAT, RAPID, and SCONE, one-to-one comparison between models is not possible. Missing values (marked using ⋯) are either because the classifier is not capable of producing these results or because the statistic was not reported. Also note that the Early Time, Bottom-level F-1 score for the ATAT classifier was not reported directly, but inferred from Figure 3(a) in G. Cabrera-Vives et al. (2024). AUC stands for area under curve.
[a] 19—way classification for ELAsTiCC using ORACLE (This work).
[b] 20—way classification for ELAsTiCC using ATAT (G. Cabrera-Vives et al. 2024).
[c] 12—way classification for ZTF simulations using RAPID (D. Muthukrishna et al. 2019).
[d] 6—way classification for PLAsTiCC using SCONE (H. Qu et al. 2021).
[e] 5—way classification between broad classes, most comparable to *level_2* for ORACLE, using a version of SUPERNNOVA (A. Möller & T. de Boissière 2020) reported by B. M. O. Fraga et al. (2024).

(G. Cabrera-Vives et al. 2024), RAPID (D. Muthukrishna et al. 2019), SCONE (H. Qu et al. 2021), and SUPERNNOVA (A. Möller & T. de Boissière 2020). ORACLE achieves performance competitive with these other general-purpose photometric classifiers at all post-trigger phases. When compared with ATAT, we observe marginally worse performance at early times, likely because the WHXE loss function causes the model to preferentially learn the top of the hierarchy (as a result of the $\alpha$ parameter described in Section 6). At late times, ORACLE improves and achieves classification performance indistinguishable from ATAT. Notably, both classifiers struggle to classify SN subtypes and CaRTs (see Section 9.2 for more details).

ORACLE also maintains comparable levels of performance against RAPID, which may be explained, in part, by the similarities in the architecture of the two models. Crucially, ORACLE does a better job at classifying core collapse SNe (such as Type Ib/c and Type II) at early epochs and maintains this advantage at later epochs (see Figure 7 in D. Muthukrishna et al. 2019).

When compared to models trained for less granular classification, like SCONE and SUPERNNOVA, ORACLE maintains similar or better performance. For example, it does a noticeable better job with classifying Fast Transients when using the full light curve when compared to the SUPERNNOVA-broad statistics reported in B. M. O. Fraga et al. (2024), while maintaining similar performance levels for all other classes.

Of course, one of the inherent strengths of ORACLE is that, unlike flat models, it can classify at many different levels of granularity in the same inference step while meeting the performance thresholds for high-throughput, real-time applications. Thus, ORACLE or future hierarchical models like it can replace a whole family of classifiers. Given the diversity in approaches and performance of photometric classifiers, further research and formal benchmarking are required to identify general trends of best approaches going forward. Such analysis is left for future work.

It is worth noting that, while ORACLE can meet or exceed the performance demonstrated by other models, it does share some of the same compromises and failure modes. We discuss these limitations in Section 9. We also note that, while general-purpose classifiers struggle with granular classifications at early times, there are several classifiers with narrow science goals that have achieved promising performance despite limited photometry. Some examples include early classification of Type Ia SNe (M. Leoni et al. 2022; A. Möller et al. 2024; N. Rehemtulla et al. 2024) and kilonovae (B. Biswas et al. 2023).

### 8.3. The Scientific Impact of Early Hierarchical Classification

In this section, we concretely demonstrate how the hierarchical classifications made possible by ORACLE can help with the earlier identification, and possibly follow-up, of rare transients such as tidal disruption events (TDEs) and KNe.

Figure 11 shows the distribution of days required for ORACLE to produce a confident classification (class score > 0.9) for TDEs in the test set. Among the TDEs that we do classify with high confidence, we can clearly see that level_1 (median of 1 day) and level_2 (median of 11 days) classifications pass the confidence threshold much sooner than the leaf (median of 21 days) classification. This allows observers to request spectroscopic follow-up with *significantly* more time for event characterization, while still knowing more about the source than would be possible with a flat multilabel classifier at these phases.

This advantage is further compounded by the fact that we can classify a larger fraction of all TDE with high confidence at level_1 and level_2 compared to the leaf. Thus, it is possible to report high-confidence classifications for sources where granular classification is difficult, which is a demonstration of our goal to align the granularity of the inference task we consider with the data we have to achieve it. Misclassification at these levels is also likely to lead to scientifically valuable data for other rare transients (as we discuss in greater detail in Section 9).

For even rarer classes of transient events, like KNe, the advantage is even more apparent. For example, ORACLE can typically produce a confident classification that the KN source is a new transient (as opposed to variable) 1 day post-trigger, while it typically takes 6 days to produce a confident classification of KN. This 5 day delay between the level_1 and leaf classification gives observers a huge advantage to narrow down their search for an electromagnetic (EM) counterpart to a gravitational wave event and collect richer data with different instruments. Considering the average discovery window expected for the EM counterpart to a binary neutron star merger is only 3–4 days during LIGO-Virgo-KAGRA's fourth observing run (V. G. Shah et al. 2024), this early classification at level_1 could be the difference between getting richer data on the KN and missing it entirely. We caution that our simulations have not included bogus alerts (from e.g., subtraction issues, cosmic rays, or other artifacts), and these remain a primary contaminant for very early classification.

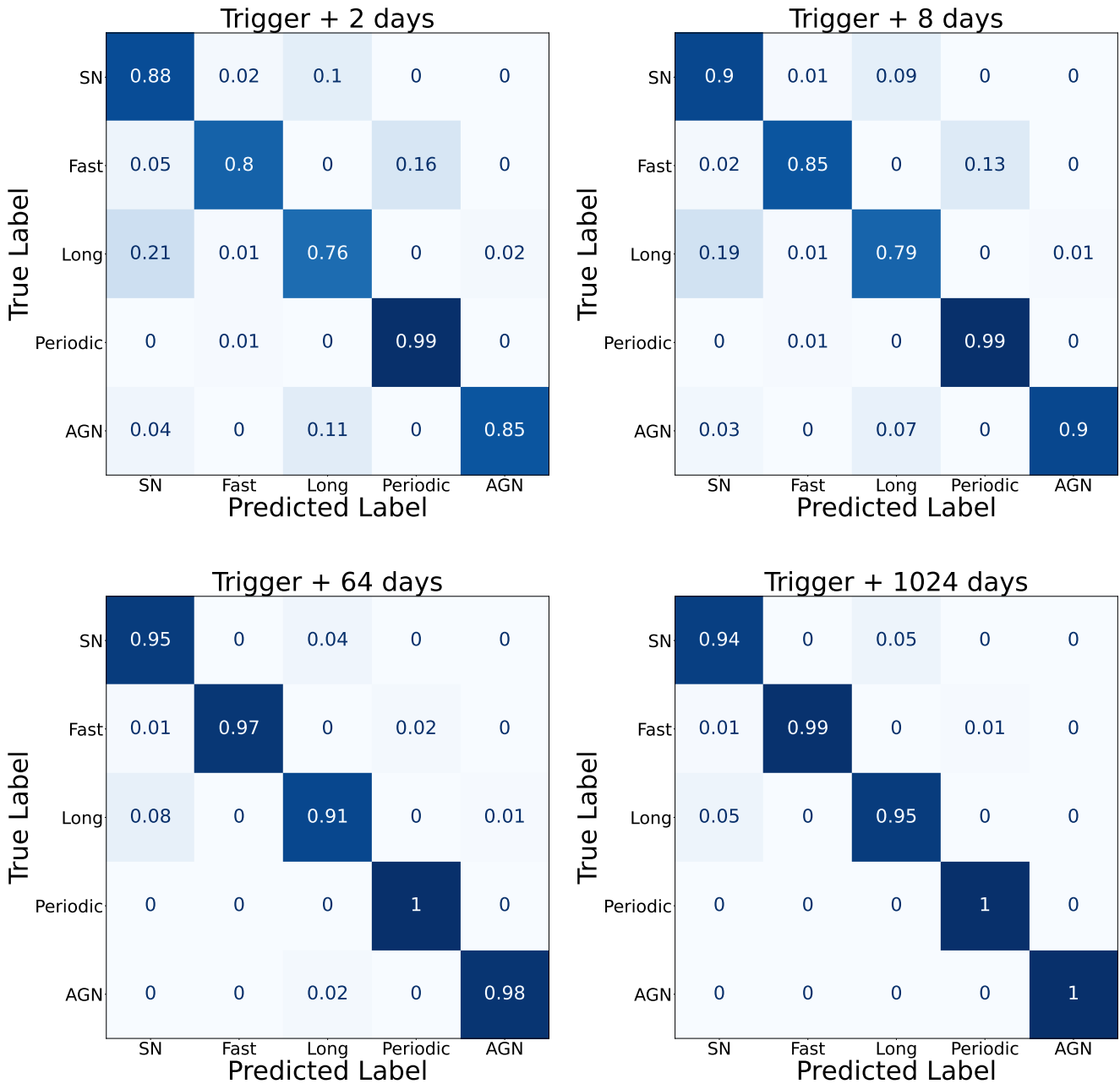

**Figure 8.** Confusion matrices for classification at level 2, as a function of the number of days since event trigger.

## 9. Discussion

While Neural Networks are often called "black box" models, we can study some of ORACLE's predictions and explore possible explanations for them. However, we must exercise caution when interpreting these results. This is especially true for classifiers trained on synthetic data, since misclassification may be reflective of data limitations at a specific phase (e.g., early classification being inherently difficult), limitations of the classifier architecture, or the limitations of the observational models used to simulate the astrophysical classes.

Nonetheless, in this section, we examine a few systematic misclassifications observed in ORACLE. We see that, in general, ORACLE "makes better mistakes" when compared to flat classifiers and misclassifies events mainly due to data limitations, similar to the predictions of a human at comparable phases.

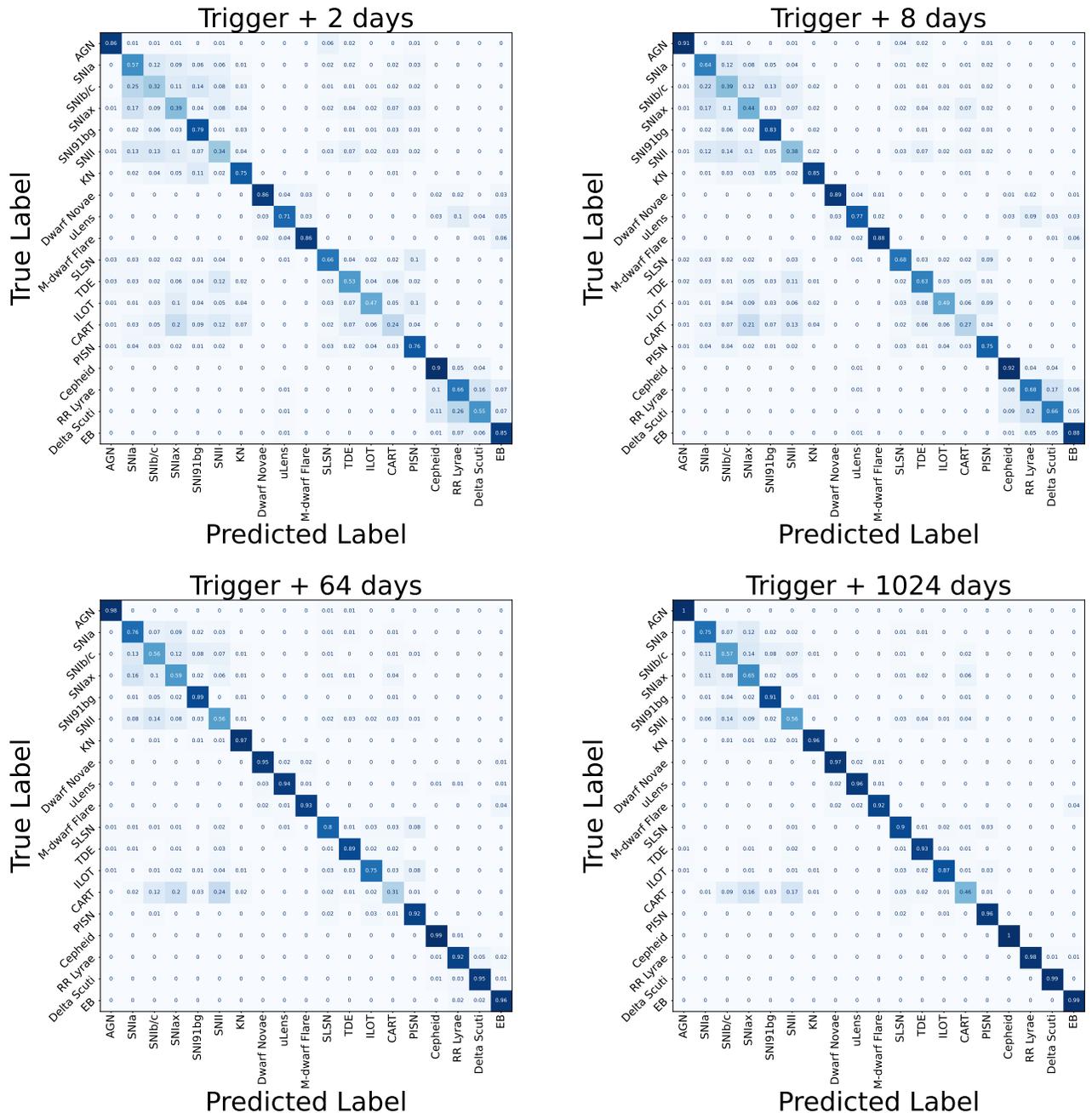

**Figure 9.** Confusion matrices for classification at the leaves, as a function of the number of days since event trigger.

### 9.1. Misclassification of Long Transients

From the confusion matrix in Figure 9, it is evident that SN subtypes act as contaminants for long transients at the leaf level. Interestingly, this corresponds to contamination between the SN and the Long classes at level_2 of the taxonomy (Figure 8)—a contamination that can also be seen in the SUPERNNOVA classifier (A. Möller & T. de Boissière 2020; B. M. O. Fraga et al. 2024).

As explained in the discussion for the loss function (Equation (2)), the model outputs the conditional probabilities for each node in the taxonomy. We then multiply the conditional probabilities along the path from the root node to the node of the class, in order to compute the true class probabilities. In doing so, we propagate probabilities downstream.

This method leads to desired behavior for correctly classified sources, since more granular predictions are strengthened by correct classifications higher in the taxonomy. When the classifier does mislabel a source higher up in the taxonomy, and that misclassification leads to leaf-level classifications, we argue that this is still the preferred behavior. We have already demonstrated that predictions higher in the taxonomy are more reliable, and thus the presence of low classification probabilities (between long transients and SNe) at level_2 should affect more granular predictions. Indeed, if the data is not sufficient to break basic degeneracies, our classifiers should not be producing confident classifications at lower levels in the taxonomy.

These choices manifest as "blocky" structures in the leaf confusion matrix at early phases (Figure 9). In particular, when ORACLE mislabels a long variable as an SN at level_2, it may then proceed to pick the highest likelihood SN as the true class of the source, resulting in this contamination for long transients.

As more data are made available and it becomes possible to break basic degeneracies (level_2), downstream (leaf) classification performance also improves.

### 9.2. Misclassification of CaRTs

Another takeaway from Figures 6 and 9 is the poor classification performance for calcium-rich transients (CaRTs) compared to other classes. CaRTs are often misclassified as SNe Ib/c, SNe-Iax, or SNe-II, even at late times. Part of this contamination may be explained by the inherent similarities in the light curves for these classes (Figure 12), especially when they are poorly sampled.

For example, the contamination with Type Iax SNe may be explained by virtue of both being fainter and redder than other SN subtypes, as well as having faster rise times. Additionally, it is not obvious how the late-time spectroscopic differences between CaRTs and SN subtypes (such as Ib, Ic, and II) are imprinted on the photometry. CaRTs and other core collapse SNe are thought to have similar progenitor systems, which may contribute to similarities in their light curves.

We note that ORACLE is not unique in these misclassifications. Other classifiers trained on ELAsTiCC and ZTF data similarly struggle to classify CaRTs (D. Muthukrishna et al. 2019; G. Cabrera-Vives et al. 2024). D. Muthukrishna et al. (2019) postulate that the contamination likely stems from the fast rise times of CaRTs, which is similar to core collapse SNe. G. Cabrera-Vives et al. (2024) report better classification performance for CaRTs using a baseline Random Forest model and find some evidence that it is due to the relatively small number of samples in the training set and the ability of a model to overfit to the training set.

Thus, while it is clear that part of the difficulty in classifying CaRTs comes from similarities in their light curves with other SN types, the relatively better performance with a Random forest model hints at the need for improved minority-class data augmentations for neural network-based methods like ORACLE (this work), ATAT, and RAPID. A more detailed analysis is required to understand if the root problem is the limited training data or something else entirely.

### 9.3. SNII and TDE Confusion

At early times, ORACLE often mislabels TDEs as Type II SNe (Figure 9). Intuitively, these classes should be easily separable based on their HOSTGAL_SNSEP parameter, which describes the angular separation between the transient and the center of their matched host galaxy in arcseconds (assuming the majority of galaxies are correctly identified). Figure 13 shows the distribution for the Transient-Host separation for both Type II SNe and TDEs. Here, we observe significant overlap between the two classes, which makes this separation difficult at early times. This problem is made worse by the fact that about 1.4% of Type II SNe and about 10% of TDEs do not have a reported HOSTGAL_SNSEP value in ELAsTiCC.

Without reliable contextual information, early transient observations are insufficient to break the degeneracy between these two classes. As more photometry becomes available, our model is able to distinguish between these events based on light curve characteristics, and we achieve better separation between the two classes.

### 10. Conclusion

In this work, we demonstrate that hierarchical models are incredibly effective at classifying astrophysical phenomena at late times, while also providing useful class information at earlier times compared to flat multilabel classifiers. This makes hierarchical classification methods particularly desirable for real-time applications in astrophysics, where an observational taxonomy exists and where full-phase light curves will not always be available.

Furthermore, the comparable performance of ORACLE relative to state-of-the-art, nonhierarchical models from the literature, at every level of the taxonomy, highlights that there may be few drawbacks to adopting this approach for future surveys. By increasing the fidelity of time-domain simulations used to train these photometric classifiers (e.g., with the injection of bogus alerts and with larger samples of rare classes), additional differences between techniques may emerge.

ORACLE models are both open-source and open-weight. All the codes for the training and analysis of the models, including a pip-installable package, are publicly available on GitHub.[14] A frozen version of the software is available on Zenodo (V. Shah et al. 2025). Additionally, we include a general-purpose Keras/Tensorflow implementation for the HXE and WHXE losses within the repository. This implementation can be easily adopted for other hierarchical classification tasks, including those outside of astronomy.

### 11. Future Work

Given the potential value of ORACLE for early follow-up of scientifically valuable targets (that can be flagged ∼10 days earlier than is possible with a leaf-level classification, as shown in Figure 11), we aim to integrate ORACLE end-to-end into automated time-domain follow-up systems for Rubin. One such example is the Recommendation System for Spectroscopic Follow-up (RESSPECT, E. E. O. Ishida et al. 2019; N. Kennamer et al. 2020; A. Wasserman et al. 2024). Whereas the initial RESSPECT framework considered cosmological constraints as the basis for spectroscopic classification of photometric SNe Ia, a modified version is planned to automatically obtain spectra of LSST events from ORACLE Level 2 classifications. LSST observations will be gradually incorporated into the ORACLE training set alongside ELAsTiCC data, and we will conduct a hyperparameter search to minimize ORACLE's computational footprint and enable regular retraining (quantization and other model compression strategies may also be possible, e.g., T. Allam et al. 2023). Several aspects of this project are already underway. We also aim to deploy ORACLE as an ANTARES (T. Matheson et al. 2021) filter to make LSST hierarchical classification results publicly available during survey operations.

Longer-term developments of ORACLE such as the ones proposed above, will require the start of Rubin operations. In particular, since the simulated ELAsTiCC data set was used to train ORACLE, it is likely that systematic errors introduced through the data synthesis process (e.g., unrealistic angular offsets) were learned, impacting the classification of real

---

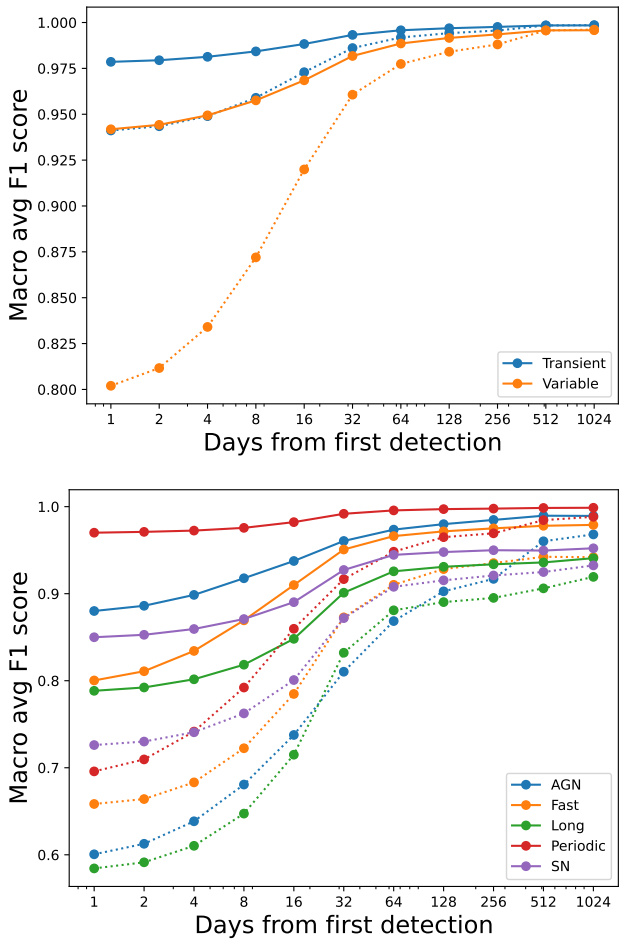

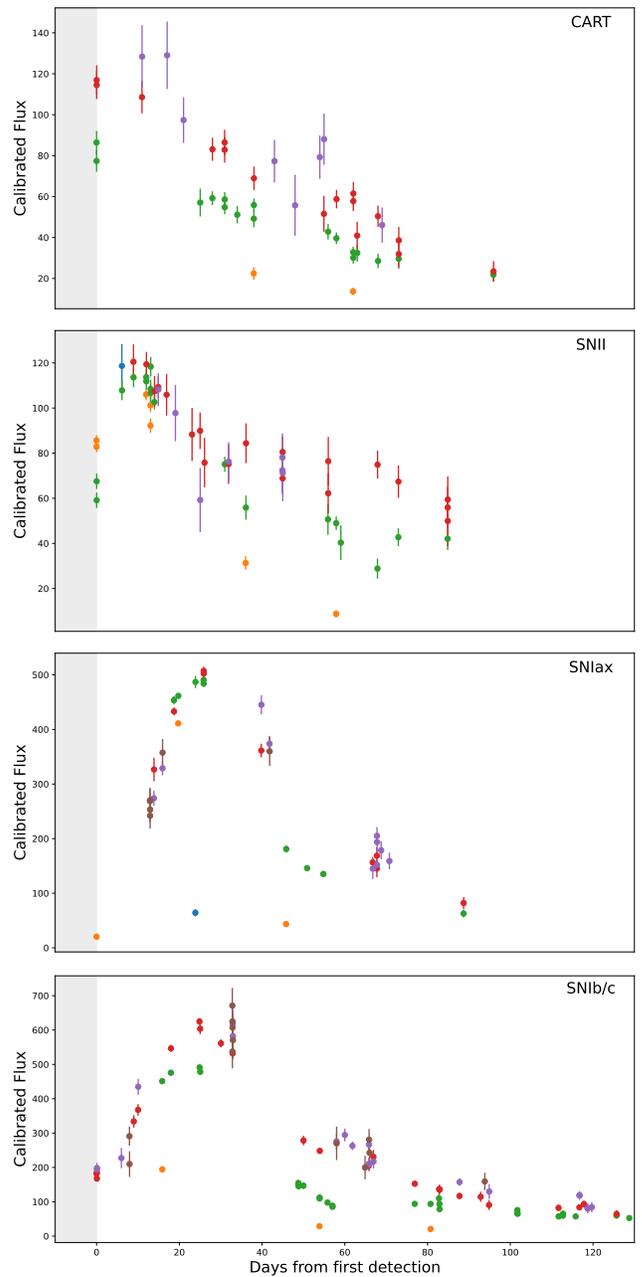

**Figure 10.** Per-class F-1 scores for the level_1 (top) and level_2 (bottom) as a function of the number of days since event trigger, for the full ORACLE model (solid lines) as well as ORACLE-lite (dotted lines).

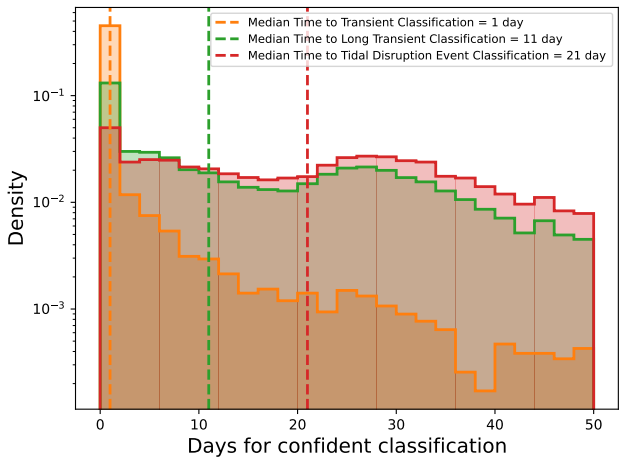

**Figure 11.** Distribution of the number of days (since trigger) required for confident classification of TDEs at the three levels of granularity. There is a clear increase in the granularity of confident classifications produced by ORACLE as more photometry becomes available at later times.

**Figure 12.** Examples of well-sampled light curves for CaRT, SNII, SNIax, and SNIb/c from the ELAsTiCC data set.

classify astrophysical phenomena at the most granular level (leaf level in our case), and thus these models also need to be trained with labels of varying granularity. Ultimately, simulated data sets such as ELAsTiCC are approximations of the time-domain sky, and fine-tuning and validation will still be needed with real LSST data to ensure reliable performance. Domain adaptation strategies show significant promise for this task.

There is scope to improve how ORACLE deals with time-independent features as well. Currently, the time-independent features assume that a galaxy has been detected and has a sufficient signal-to-noise ratio to be processed by the Rubin pipeline to infer brightness. This photometry is also necessary to derive host properties such as the photometric redshift. This could potentially introduce a bias against sources with very

Rubin observations. Additionally, even if errors introduced by data synthesis could be avoided, ELAsTiCC is incomplete. There are entire classes of astrophysical transients that are not represented, and beyond these known classes, several astrophysical transients that have been predicted but have not been observed to date. Moreover, it is not always possible to low surface brightness (or whose luminosity evolves on timescales greater than the duration between the reference template and a nightly observation). A different approach is to use a deep coadded postage stamp of some fixed-size region around the transient. This will allow the network to classify both very low-redshift transients (where the host galaxy is very extended and host-galaxy magnitudes are ill-defined) as well as transients in very low-surface-brightness galaxies.

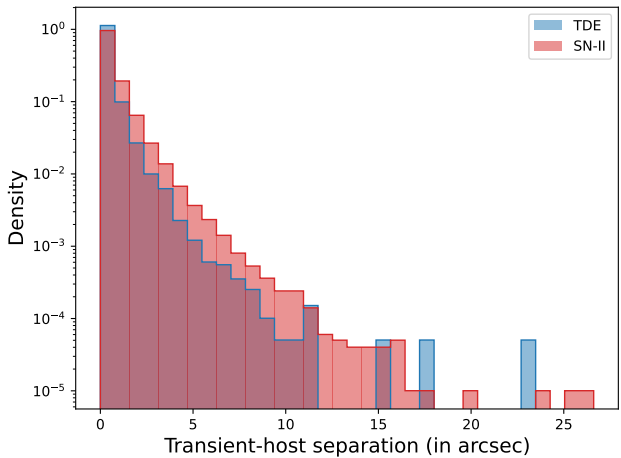

**Figure 13.** Transient-host separation for Type II SNe and TDEs.

Finally, approaches to calibrate NN outputs to reflect true probabilities for astrophysical applications have been demonstrated in recent works (A. Möller & T. de Boissière 2020). Given the downstream benefits of calibration, we would be interested in calibrating ORACLE's output pseudoprobability scores. However, doing so in the context of a hierarchical classifier remains largely unexplored and is left for future work.

## Acknowledgments

V.S. acknowledges the support of the LSST Corporation's 2021 Enabling Science award, the 2024 Stanley Wyatt Award, and the 2024 Preble Scholarship for making this work possible. This work made use of the Illinois Campus Cluster, a computing resource that is operated by the Illinois Campus Cluster Program (ICCP) in conjunction with the National Center for Supercomputing Applications (NCSA) and which is supported by funds from the University of Illinois at Urbana-Champaign. This work was partially supported by the Center for AstroPhysical Surveys (CAPS) at the National Center for Supercomputing Applications (NCSA), University of Illinois Urbana-Champaign. V.S. also acknowledges computational resources provided by the Pittsburgh Supercomputing Center (Bridges-2; S. T. Brown et al. 2021), which were used for initial prototyping.

G.N. gratefully acknowledges NSF support from AST-2206195, and a CAREER grant AST-2239364, supported in part by funding from Charles Simonyi, and OAC-2311355, DOE support through the Department of Physics at the University of Illinois, Urbana-Champaign (#13771275), and support from the HST Guest Observer Program through HST-GO-16764, and HST-GO-17128 (PI: R. Foley). This work was performed in part at the Aspen Center for Physics, which is supported by National Science Foundation grant PHY-2210452. Support was provided by Schmidt Sciences, LLC., for K.M. A.I.M. is supported by Schmidt Sciences.

This work is supported by the National Science Foundation under Cooperative Agreement PHY-2019786 (The NSF AI Institute for Artificial Intelligence and Fundamental Interactions, http://iaifi.org/).

We gratefully acknowledge the support of the NSF-Simons AI Institute for the Sky (SkAI) via grants NSF AST-2421845 and Simons Foundation MPS-AI-00010513.

Finally, this paper has undergone internal review in the LSST Dark Energy Science Collaboration (DESC). The DESC acknowledges ongoing support from the Institut National de Physique Nucléaire et de Physique des Particules in France; the Science & Technology Facilities Council in the United Kingdom; and the Department of Energy, the National Science Foundation, and the LSST Corporation in the United States. DESC uses resources of the IN2P3 Computing Center (CC-IN2P3–Lyon/Villeurbanne—France) funded by the Centre National de la Recherche Scientifique; the National Energy Research Scientific Computing Center, a DOE Office of Science User Facility supported by the Office of Science of the U.S. Department of Energy under Contract No. DE-AC02-05CH11231; STFC DiRAC HPC Facilities, funded by UK BEIS National E-infrastructure capital grants; and the UK particle physics grid, supported by the GridPP Collaboration. This work was performed in part under DOE Contract DE-AC02-76SF00515. The authors want to extend their gratitude to Anais Möller, who served as the DESC Internal Reviewer; Maria Vincenzi and Dillon Brout, who served as the Time-Domain Working Group Conveners; and Douglas Clowe, who served as the Publication Manager on behalf of the LSST Dark Energy Science Collaboration, and Nabeel Rehemtulla for their invaluable feedback and guidance on this work.

Author contributions are listed below:

1. V. G. Shah: Software, writing, and editing.
2. A. Gagliano: Software, writing, and editing.
3. K. Malanchev: Oversight, writing, and editing.
4. G. Narayan: Oversight, writing, and editing.
5. A. Malz: Simulations.

*Data Note.* We want to acknowledge the contributions of the team that created the ELAsTiCC data set: Gautham Narayan, Alex Gagliano, Alex Malz, Catarina Alves, Deep Chatterjee, Emille Ishida, Heather Kelly, John Franklin Crenshaw, Konstantin Malanchev, Laura Salo, Maria Vincenzi, Martine Lokken, Qifeng Cheng, Rahul Biswas, Renée Holžek, Rick Kessler, Robert Knop, Ved Shah Gautam.

*Software Note.* This work makes use of NUMPY (C. R. Harris et al. 2020), ASTROPY (Astropy Collaboration et al. 2013, 2018, 2022), SCIPY (P. Virtanen et al. 2020), MATPLOTLIB (J. D. Hunter 2007), PANDAS (W. McKinney 2010), NETWORKX (A. Hagberg et al. 2008), TENSORFLOW (M. Abadi et al. 2015), and SKLEARN (F. Pedregosa et al. 2011).

## Appendix A
## SNANA Model Mappings

Mappings from SNANA models to astrophysical classes (Table 9).

**Table 9**
Mapping between the SNANA Model and the Astrophysical Class of the Object

| SNANA Model | Astrophysical Class | Description |
|---|---|---|
| SNII-NMF | SNII | Type II—Supernova |
| SNIc-Templates | SNIb/c | Type Ic—Supernova |
| CaRT | CaRT | CaRTs |
| EB | EB | Eclipsing Binary |
| SNIc+HostXT_V19 | SNIb/c | Type Ic—Supernova |
| d-Sct | Delta Scuti | Delta Scuti |
| SNIb-Templates | SNIb/c | Type Ib—Supernova |
| SNIIb+HostXT_V19 | SNII | Type II—Supernova |
| SNIcBL+HostXT_V19 | SNIb/c | Type Ic—Supernova |
| CLAGN | AGN | Changing Look AGN |
| PISN | PISN | Pair Instability Supernova |
| Cepheid | Cepheid | Cepheid Variable Star |
| TDE | TDE | TDE |
| SNIa-91bg | SNI91bg | Type Ia—Supernova, 91bg-like |
| SLSN-I+host | SLSN | Superluminous Supernova |
| SNIIn-MOSFIT | SNII | Type IIn—Supernova |
| SNII+HostXT_V19 | SNII | Type II—Supernova |
| SLSN-I_no_host | SLSN | Superluminous Supernova |
| SNII-Templates | SNII | Type II—Supernova |
| SNIax | SNIax | Type Iax—Supernova |
| SNIa-SALT3 | SNIa | Type Ia—Supernova |
| KN_K17 | KN | KN |
| SNIIn+HostXT_V19 | SNII | Type IIn—Supernova |
| dwarf-nova | Dwarf Novae | Dwarf Novae |
| uLens-Binary | uLens | Microlensing Event, Binary Lens |
| RRL | RR Lyrae | RR Lyrae Variable |
| Mdwarf-flare | M Dwarf Flare | M Dwarf Flare |
| ILOT | ILOT | Intermediate-luminosity Optical Transient |
| KN_B19 | KN | KN |
| uLens-Single-GenLens | uLens | Microlensing Event, Single Lens |
| SNIb+HostXT_V19 | SNIb/c | Type Ib—Supernova |
| uLens-Single_PyLIMA | uLens | Microlensing Event, Single Lens |

## Appendix B
## ROC Curves

Figures 14, 15, and 16 show the receiver operating characteristic (ROC) curves for the ORACLE model at different phases of light-curve evolution.

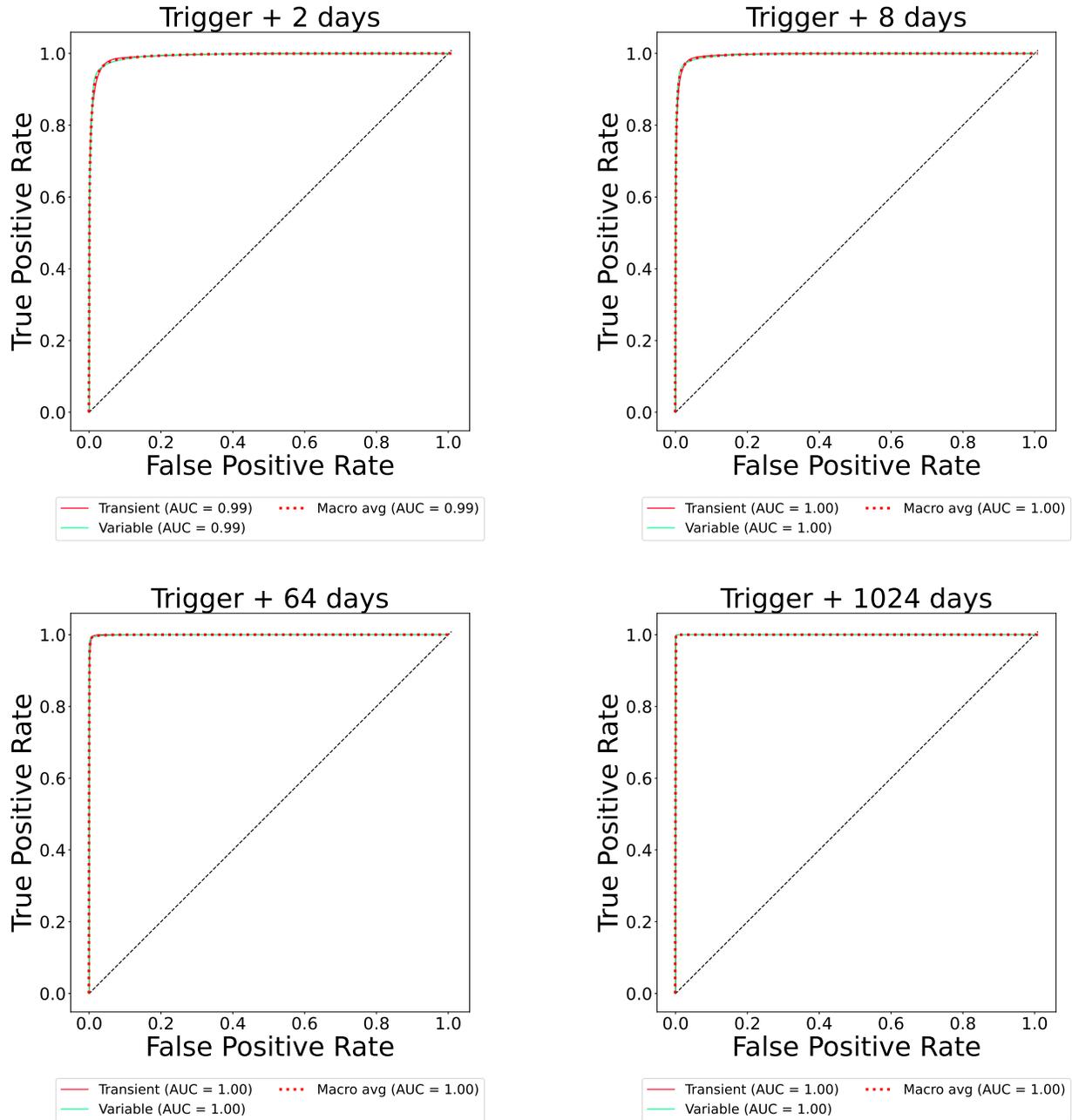

**Figure 14.** ROC curves for classification at level 1, as a function of the number of days since the event trigger.

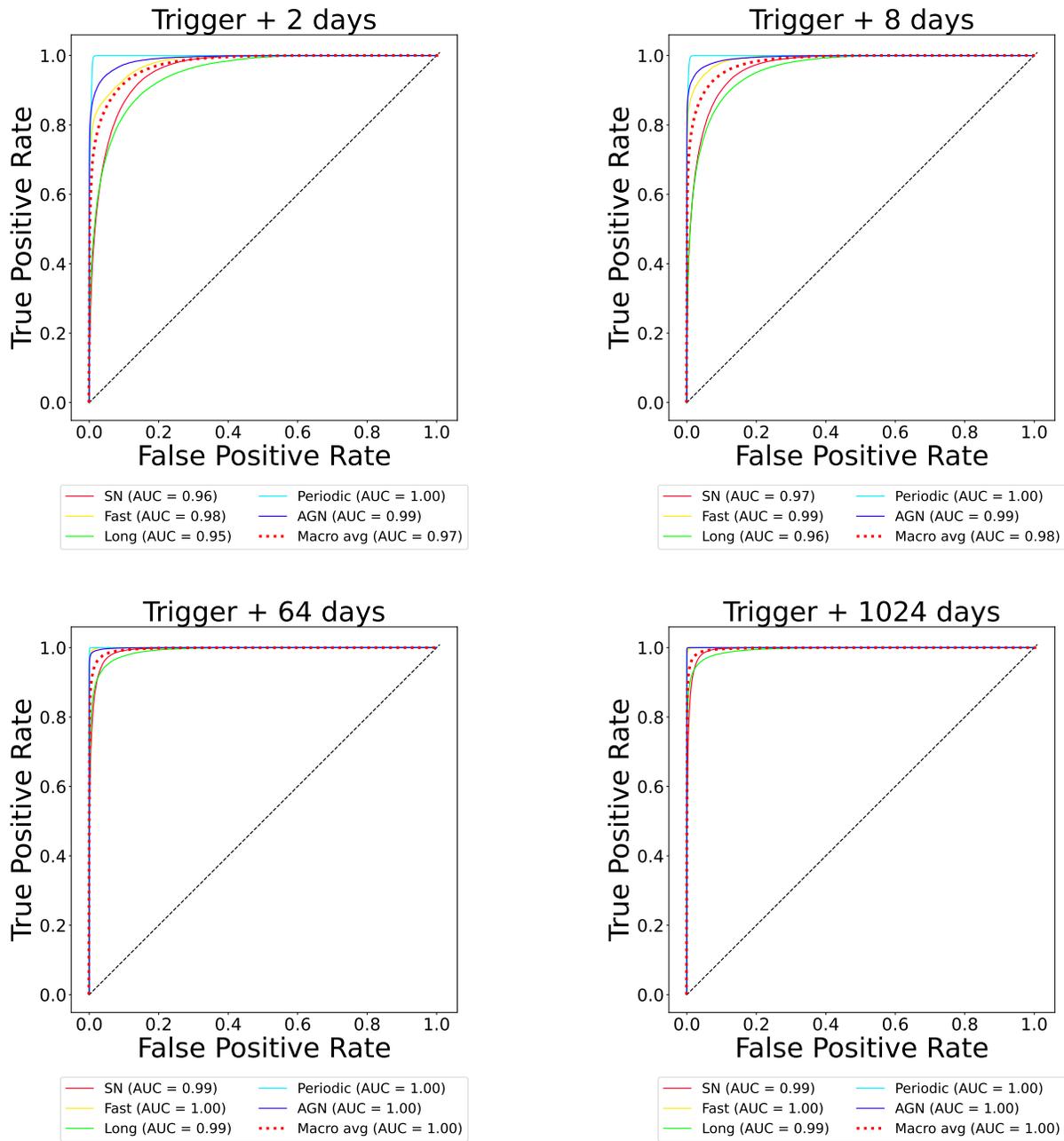

**Figure 15.** ROC curves for classification at level 2, as a function of the number of days since the event trigger.

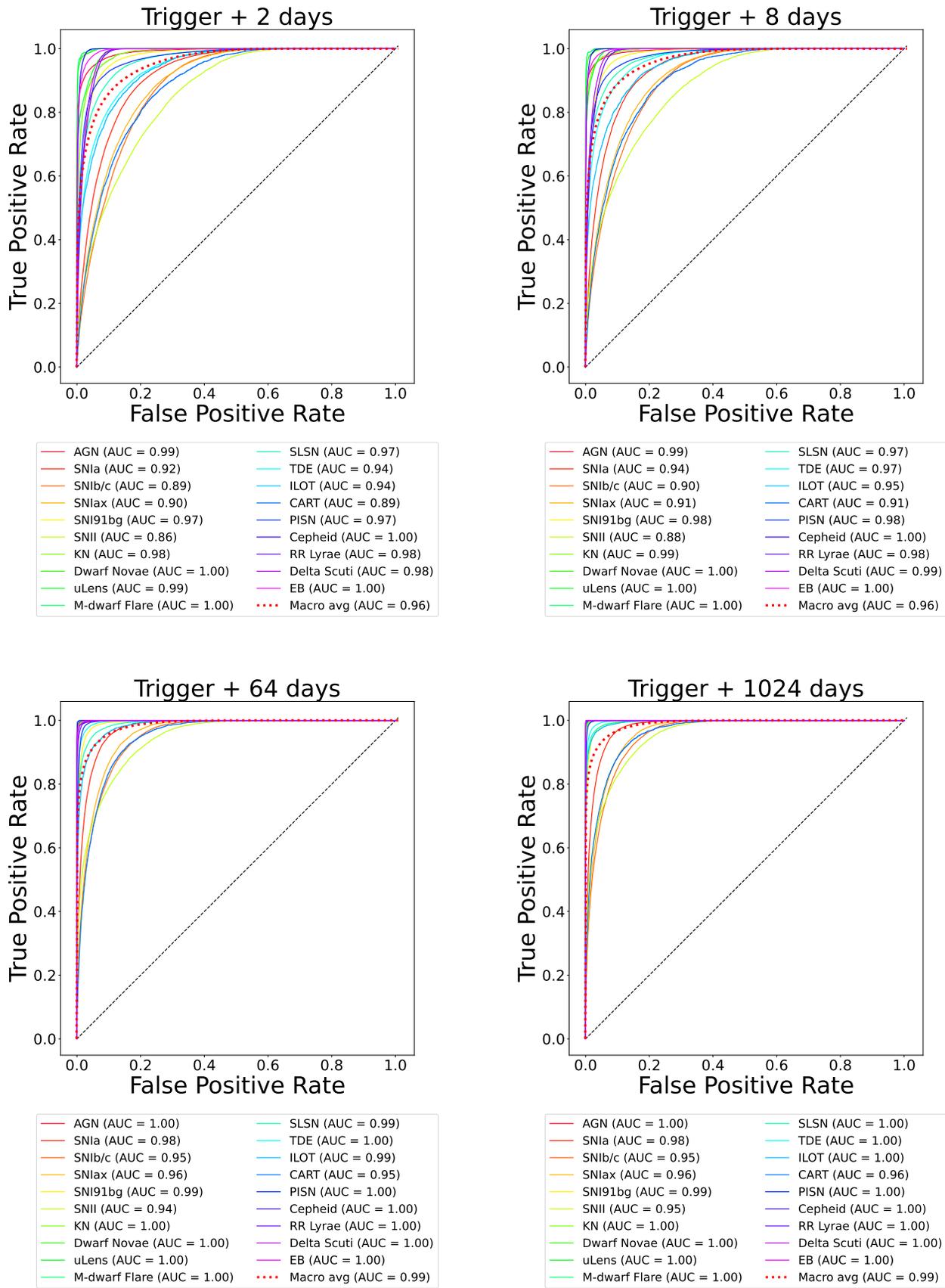

**Figure 16.** ROC curves for classification at the leaves, as a function of the number of days since the event trigger.

## Appendix C
## Class Scores as a Function of Days Since Trigger

Figures 17 and 18 show the evolution of the mean class scores for all leaf classes in the taxonomy as a function of days since trigger, organized by the true class.

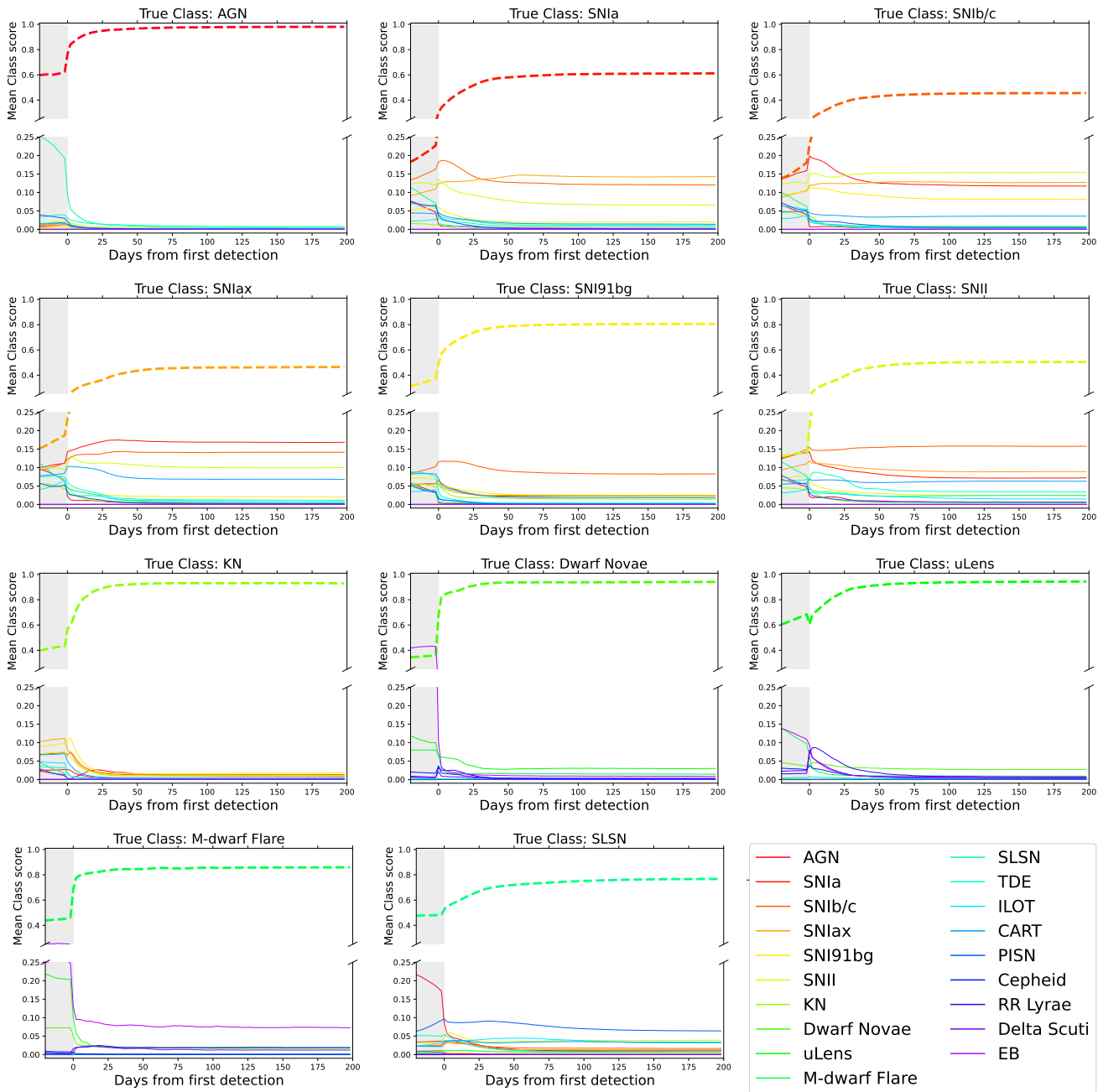

**Figure 17.** Mean class scores (across all samples of the true class) for all leaf classes in the taxonomy as a function of days since trigger, categorized by the true class. The wide dashed line in each plot shows the mean class score (or pseudoprobability) of the true class. The shaded gray region highlights the pretrigger duration. Continued in Figure 18.

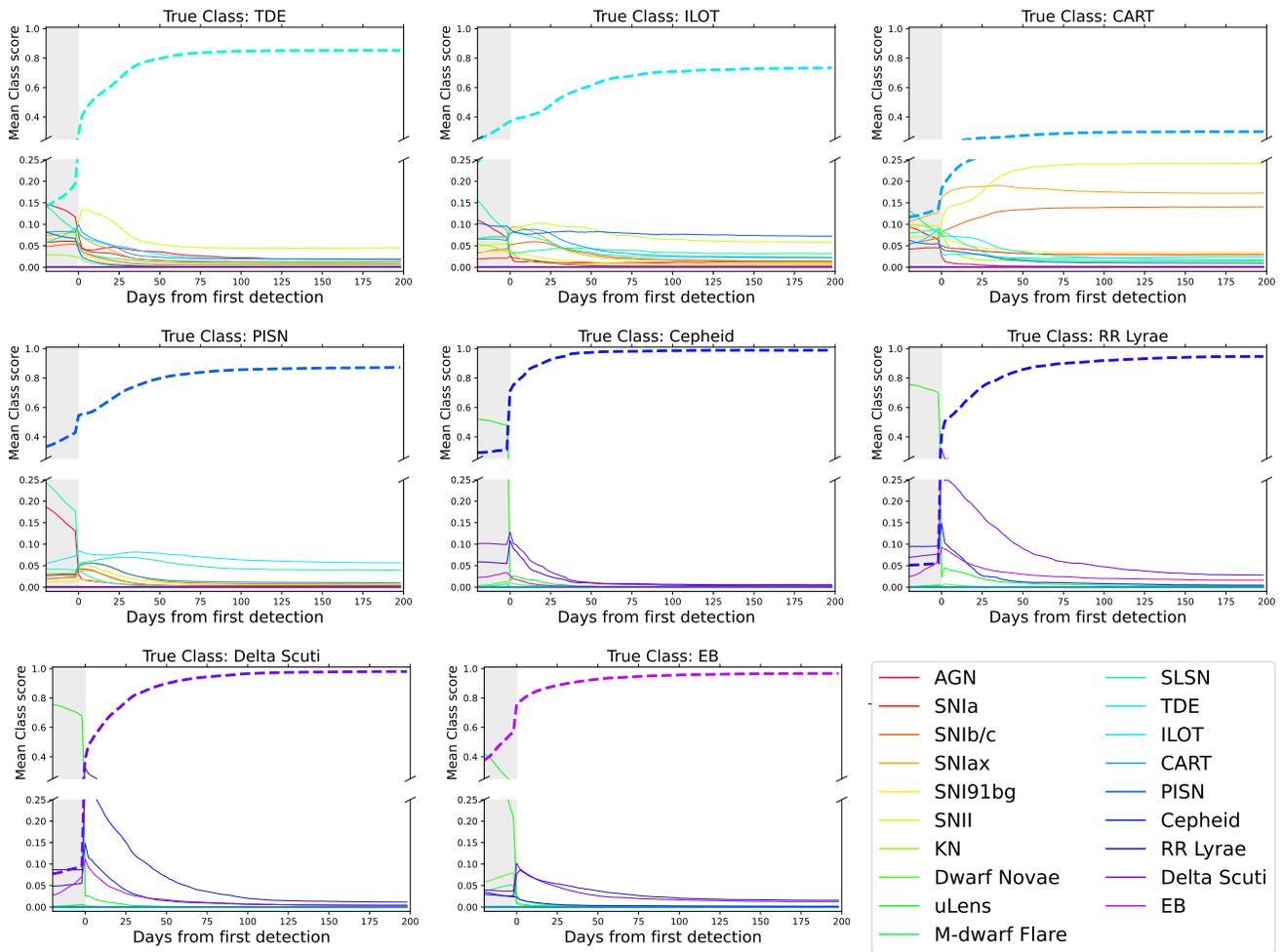

**Figure 18.** Continuation of data displayed in Figure 17.

## Appendix D
## Sample Light Curves

Figures 19 and 20 show example light curves for each class in the data set, along with the time evolving true class score, as generated by ORACLE.

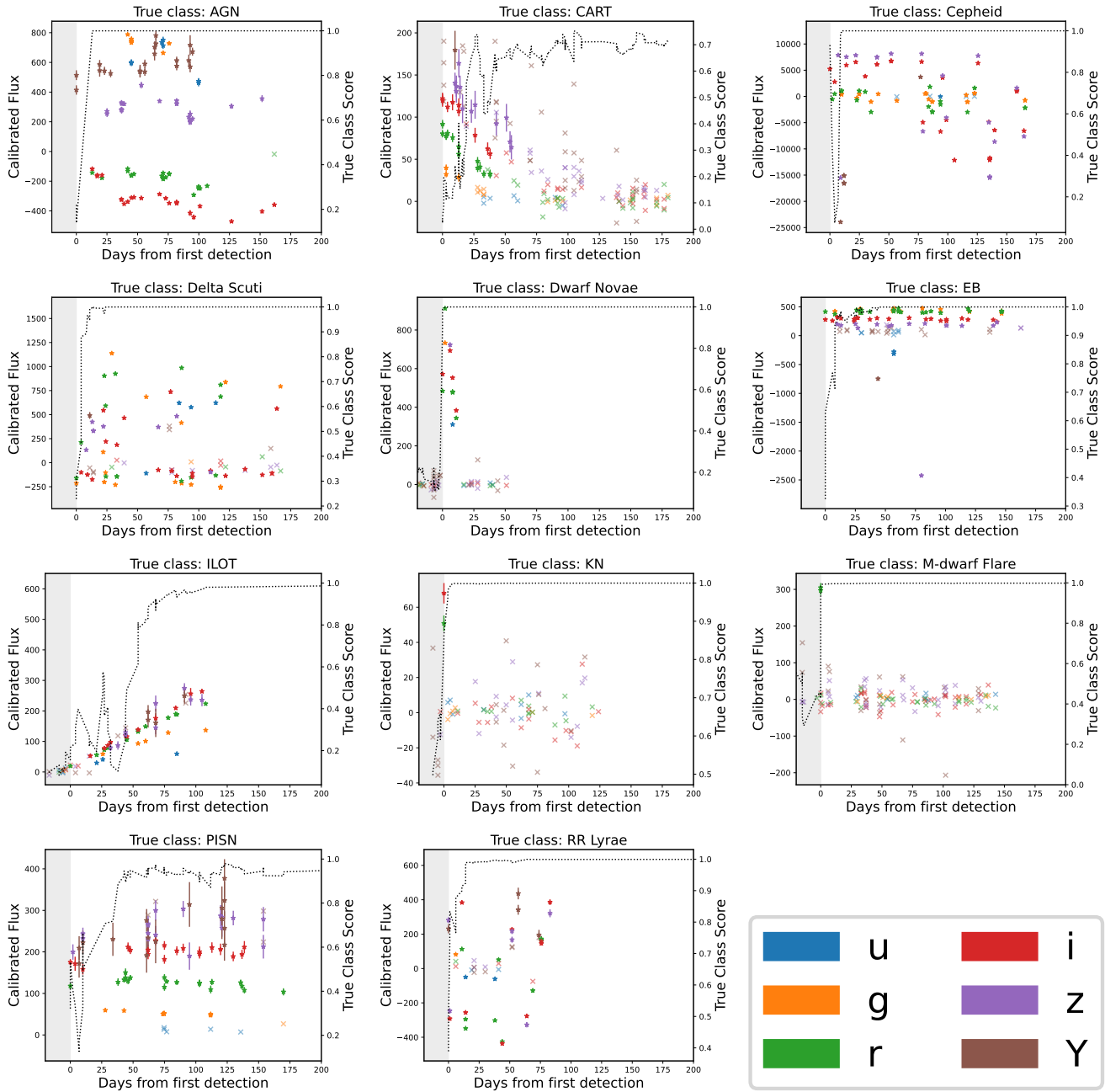

**Figure 19.** Sample light curves, using SNANA calibrated flux (see Table 1, Equation (1)), from each of the 19 leaf classes. The predicted class score of the true class is plotted using a dotted black line. The detections are plotted using the star marker with error bars, while nondetections are plotted using the cross marker. Continued in Figure 20.

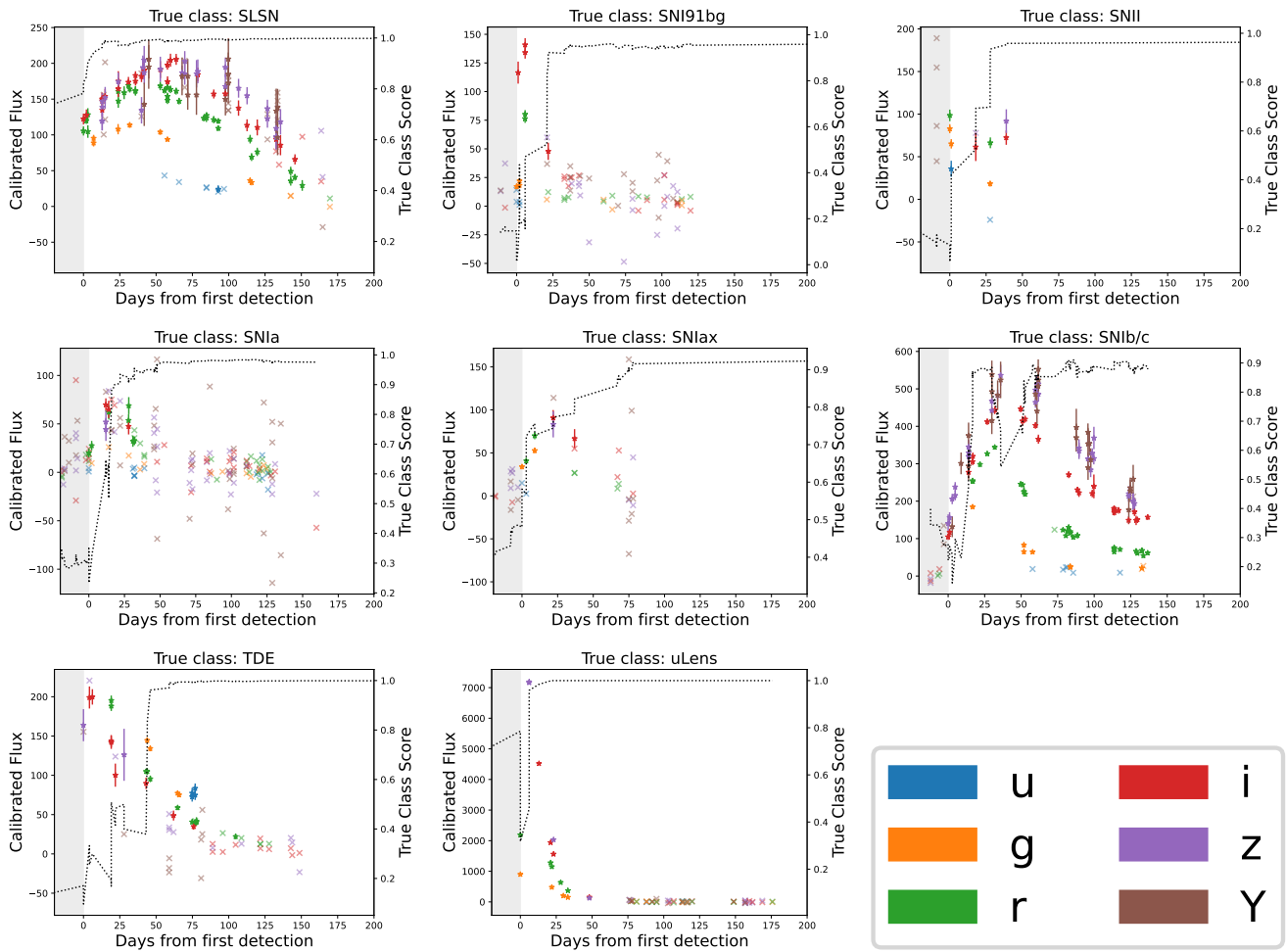

**Figure 20.** Continuation of data displayed in Figure 19.

## ORCID iDs

Ved G. Shah  https://orcid.org/0009-0009-1590-2318
Alex Gagliano  https://orcid.org/0000-0003-4906-8447
Konstantin Malanchev  https://orcid.org/0000-0001-7179-7406
Gautham Narayan  https://orcid.org/0000-0001-6022-0484
Alex I. Malz  https://orcid.org/0000-0002-8676-1622

## References

Abadi, M., Agarwal, A., Barham, P., et al. 2015, TensorFlow: Large-scale Machine Learning on Heterogeneous Systems, https://www.tensorflow.org/
Aleo, P. D., Engel, A. W., Narayan, G., et al. 2024, ApJ, 974, 172
Allam, T., Jr., Peloton, J., & McEwen, J. D. 2023, arXiv:2303.08951
Astropy Collaboration, Price-Whelan, A. M., Lim, P. L., et al. 2022, ApJ, 935, 167
Astropy Collaboration, Price-Whelan, A. M., Sipőcz, B. M., et al. 2018, AJ, 156, 123
Astropy Collaboration, Robitaille, T. P., Tollerud, E. J., et al. 2013, A&A, 558, A33
Bertinetto, L., Mueller, R., Tertikas, K., Samangooei, S., & Lord, N. A. 2019, arXiv:1912.09393
Biswas, B., Ishida, E. E. O., Peloton, J., et al. 2023, A&A, 677, A77
Boone, K. 2019, AJ, 158, 257
Brown, S. T., Buitrago, P., Hanna, E., et al. 2021, Practice and Experience in Advanced Research Computing 2021: Evolution Across all Dimensions, PEARC '21 (New York: ACM)
Cabrera-Vives, G., Moreno-Cartagena, D., Astorga, N., et al. 2024, A&A, 689, A289
Carrasco-Davis, R., Reyes, E., Valenzuela, C., et al. 2021, AJ, 162, 231
Farias, D., Gall, C., Narayan, G., et al. 2024, A&A, 977, 152
Foley, R. J., Challis, P. J., Chornock, R., et al. 2013, ApJ, 767, 57
Fraga, B. M. O., Bom, C. R., Santos, A., et al. 2024, A&A, 692, A208
Gagliano, A., Contardo, G., Foreman-Mackey, D., Malz, A. I., & Aleo, P. D. 2023, ApJ, 954, 6
Gagliano, A., Narayan, G., Engel, A., Carrasco Kind, M. & LSST Dark Energy Science Collaboration 2021, ApJ, 908, 170
Graham, M. J., Kulkarni, S. R., Bellm, E. C., et al. 2019, PASP, 131, 078001
Gupta, R., Muthukrishna, D., & Lochner, M. 2025, RASTI, 4, rzae054
Hagberg, A., Swart, P. J., & Schult, D. A. 2008, in Proc. 7th Python in Science Conf., ed. G. Varoquaux, T. Vaught, & J. Millman, 11
Harris, C. R., Millman, K. J., van der Walt, S. J., et al. 2020, Natur, 585, 357
Hunter, J. D. 2007, CSE, 9, 90
Ishida, E. E. O., Beck, R., González-Gaitán, S., et al. 2019, MNRAS, 483, 2
Ivezić, Ž., Kahn, S. M., Tyson, J. A., et al. 2019, ApJ, 873, 111
Kennamer, N., Ishida, E. E. O., Gonzalez-Gaitan, S., et al. 2020, arXiv:2010.05941
Kessler, R., Bernstein, J. P., Cinabro, D., et al. 2009, PASP, 121, 1028
Klimczak, H., Oszkiewicz, D., Carry, B., et al. 2022, A&A, 667, A10
Leoni, M., Ishida, E. E. O., Peloton, J., & Möller, A. 2022, A&A, 663, A13
Lochner, M., McEwen, J. D., Peiris, H. V., Lahav, O., & Winter, M. K. 2016, ApJS, 225, 31
Matheson, T., Stubens, C., Wolf, N., et al. 2021, AJ, 161, 107
McKinney, W. 2010, in Proc. 9th Python in Science Conf., ed. S. van der Walt & J. Millman, 56
Möller, A., & de Boissière, T. 2020, MNRAS, 491, 4277
Möller, A., Wiseman, P., Smith, M., et al. 2024, MNRAS, 533, 2073
Muthukrishna, D., Narayan, G., Mandel, K. S., Biswas, R., & Hložek, R. 2019, PASP, 131, 118002
Narayan, G. & ELAsTiCC Team 2023, AAS Meeting, 241, 117.01
Pedregosa, F., Varoquaux, G., Gramfort, A., et al. 2011, Journal of Machine Learning Research, 12, 2825

Qu, H., Sako, M., Möller, A., & Doux, C. 2021, AJ, 162, 67

Rehemtulla, N., Miller, A. A., Jegou Du Laz, T., et al. 2024, ApJ, 972, 7

Sánchez-Sáez, P., Reyes, I., Valenzuela, C., et al. 2021, AJ, 161, 141

Schuurmans, J., & Frasincar, F. 2023, arXiv:2308.01210

Shah, V., Gagliano, A., Malanchev, K., et al. 2025, ORACLE- A Family of
    Real-time, Hierarchical Classifiers for Transient and Variable Phenomena
    in LSST Alert Streams, v2, Zenodo, doi:10.5281/zenodo.15328166

Shah, V. G., Narayan, G., Perkins, H. M. L., et al. 2024, MNRAS, 528, 1109

Shahroudnejad, A. 2021, arXiv:2102.01792

Silverman, J. M., Nugent, P. E., Gal-Yam, A., et al. 2013, ApJS, 207, 3

Villar, V. A., Cranmer, M., Berger, E., et al. 2021, ApJS, 255, 24

Villar, V. A., de Soto, K., & Gagliano, A. 2023, arXiv:2312.02266

Virtanen, P., Gommers, R., Oliphant, T. E., et al. 2020, NatMe,
    17, 261

Wasserman, A., Cohen-Tanugi, J., Dai, M., et al. 2024, AAS Meeting, 243,
    232.02